\overfullrule=0mm
\magnification=1200
\hsize=6.0 true in
\baselineskip=12pt

\noindent{{\sl Keywords:} \quad Bilayer membranes; Dynamics; 
Hydrodynamic theory; Isolated membranes}

\bigskip
\leftline{\bf 1. Introduction}
\medskip

Phenomenological studies of lipid bilayer membranes have become 
imperative, for a large amount of experiments reveal that the matrix of 
biomembrane plays an important role in some non-linear phenomena such as 
shape fluctuations of cells and oscillations of the membrane potential in an 
external electric field (1-3). In approaching the problem, however, one is 
immediately confronted with complexities because of the concurrence of the 
elastic-viscous motion and the change in shape of the material. The lack of 
a theoretical framework tailored for non-Newtonian fluid membranes impedes 
progress in this research area, though the general geometrical method in 
continuum mechanics has been well established. The purpose of this paper is 
to present such a framework.

The membranes constituted by two layers of rod-shaped molecules are 
identified with hydrotropic liquid crystals in the laminar phase. From this 
point of view, one assumes in the present paper that the dynamic theory of 
liquid crystals developed during the last 30 years is applicable to the 
membranes. In fact, the experimental observations mentioned above would not 
be surprising if one thinks of the anisotropy feature of the material, which, 
as is well known, leads to instabilities in the liquid crystal bulk.

Due to the pioneering research of Oseen [4], the fundamental contribution 
of Frank [5] and the important supplementary work of Nehring and Saupe [6], a 
curvature elasticity theory of liquid crystals has been established. It is 
essentially a phenomenological theory. A dimensionless unit vector named 
``director" is defined to indicate the preferred orientation of the 
molecules. The energy stored in the material during elastic deformation is 
thus 
expressed in terms of this vector. Extending the theory to thin films in 
particular yields the well-known curvature elasticity equation of membranes 
[7] 
which has been used in fruitful studies on the hydrostatic form of lipid 
vesicles [8].

The dynamic theory of liquid crystals [9-13], mainly uniaxial crystals, 
stems from such a mechanical base. It assumes rotation of the 
molecules, indicated by the director, superposed on the mean flow 
of the molecules themselves. The theory was used to investigate 
instabilities of liquid crystals and has interpreted successfully the 
Williams' domain structure and the homoetropic textures as well 
in nematics [14-22].

One adapts the theory to the particular context by neglecting 
non-uniformities across the membrane thickness and replacing the 
three-dimensional film by a two-dimensional surface. It is exactly 
the method used in the theory of thin shells [23]. The simplification 
is acceptable in most cases of biocells, for the membrane thickness 
(40-80 $ \AA $) is sufficiently small with regard to the cell diameter 
(1-10 $ \mu $m). Moreover, the tilt angle of the long molecular axes 
with respect to the vector normal to the surface is supposed to be 
small and hence the elasticity theory is restricted within the extent 
of linear approximation. With these considerations, one obtains a 
particular version of the dynamic equations, in which geometric 
invariants such as the fundamental magnitudes and the curvatures 
are used to describe the shape of the membrane, while the displacement 
of the configuration of the membrane is treated separately from
the internal movement of the particles inside the membrane. The 
attempt to construct such a group of equations, so far as one knows, 
was made 10 years ago by Peterson [24]. He established, however, 
only the dynamic equations of a solid film and the hydrodynamic 
equations of a Newtonian fluid membrane, separately. This left a 
gap between these two limiting cases. One will see that by 
introducing the theory of liquid crystals the gap is naturally bridged.

Recent tether experiments support the opinion that slip between 
the two molecular layers causes interlayer viscous dissipation [25] 
which might not be negligible in dynamical studies of bilayer 
membranes [26]. One recognizes that this kind of internal dissipation 
is excluded from the present theoretical model.

This paper is the first part of a series devoted to present the 
general electrohydrodynamic equations of polar viscoelastic fluid 
membranes isolated from their surroundings. The hydrodynamic theory of 
liquid crystals will be recapitulated in the next section, then it 
will be adapted to the context of bilayer membranes in the third 
section. The connection of such a membrane to the liquid surroundings 
will be considered in subsequent papers.

\bigskip
\leftline{\bf 2. Outline of the Ericksen-Leslie theory}
\medskip

The hydrodynamics of liquid crystals was primarily established by 
Ericksen [9]. He assumed an intrinsic motion of the non-spherical 
molecules and used a time-dependent director to indicate it. Because 
of the material symmetry, he constructed the constitutive 
relationships of the dynamic variables involved and hence interpreted 
explicitly the dynamic conservation laws for uniaxial liquid crystals. 
Leslie [10] was able to take into account the antisymmetric stress of 
the anisotropic continuum [27,28], which had been overlooked in the 
expressions of Ericksen, and thus provided us a refined version of 
the theory.

\vfill\eject

\medskip
\leftline{\sl 2.1. Mass balance}
\smallskip

The principle of mass conservation holds in the convected volume $ R $
$${d \over dt} \int\!\!\!\int\limits_R\!\!\!\int \rho \, dR = 0 
\eqno (1) $$ 
where $\rho$ is the mass density and {\it t} is the time.

\medskip    
\leftline{\sl 2.2. Conservation of momentum}
\smallskip

Let {\bf v} denote the barycenter velocity of the molecules. 
Conservation of linear momentum in the motion of the continuum is 
assumed, which asserts
$$ {d \over dt} \int\!\!\!\int\limits_R\!\!\!\int \rho {\bf v} \, dR = 
\int\!\!\!\int\limits_R\!\!\!\int {\bf F} \, dR + 
\int\limits_s\!\!\!\int d {\bf s} \cdot {\bf T} \eqno (2) $$
where $ \bf F $ is the body force, $ \bf T $ the stress tensor 
and $ s $ is the surface enclosing the considered volume.
 
The body force consists of the gravitational attraction and 
electromagnetic forces [9] 
$$ {\bf F} = {\bf F}^g+ {\bf F}^{e,m} \eqno (3) $$

The gravitational force is naturally written as
$$ {\bf F}^g = \rho {\bf\hat g} \eqno (4) $$
where $ {\bf\hat g} $ is the gravitational acceleration.

The force generated in an external magnetic field is given as [9] 
$$ {\bf F}^m = \left( \bf M \cdot \bf\nabla \right) \bf H \eqno (5) $$
where $ \bf H $ is the external magnetic field, $ \bf M $ is the 
magnetization and $ \bf\nabla $ is the vector derivative operator. 

The most general relationship of $ \bf M $ linear with the magnetic field 
is [9]
$$ {\bf M} = \chi_\perp {\bf H} + \left( \chi_\parallel - 
\chi_\perp \right) \left( {\bf d} \cdot {\bf H} \right) {\bf d} = 
\chi_\perp {\bf H} + \chi_a \left( {\bf d} \cdot {\bf H} \right) 
{\bf d} \eqno (6) $$
where $ \bf d $ is the director, $ \chi_\parallel $ and 
$ \chi_\perp $ are the constant magnetic susceptiblities in the 
directions parallel and perpendicular to the director, respectively, 
and $ \chi_a $ is the anisotropy.

The force generated by the electric field is not expressed explicitly
in the Ericksen-Leslie theory. In electrohydrodynamic studies of 
liquid crystals, however, the Lorentz force is commonly considered as 
a cause of instabilities [17-22]. We do not go far towards calculating 
completely the force in the electric field, but, inaccurately, equate 
it to the Lorentz force, which amounts partly to the ponderomotive 
force in liquid dielectrics [29]. That is
$$ {\bf F}^e = \rho_e \bf E \eqno (7) $$
where $ \rho_e $ is the density of injected electric charges and 
$ \bf E $ is the electric field.

The stress tensor consists of three parts [9]
$$ {\bf T} = {\bf T}^i + {\bf T}^e + {\bf T}^v \eqno (8) $$
where $ {\bf T}^i $ is the isotropic part of the stress, 
$ {\bf T}^e $ is the stress arising from elastic deformation and 
$ {\bf T}^v $ is that from viscous motion.

The isotropic part is expressed as
$$ {\bf T}^i = -\sigma \bf I \eqno (9) $$ 
where $ \sigma $ is a constant and $ \bf I $ is the unit tensor.

The elastic stress may be deduced from the density of reversible 
deformation work $ W^d $ [9], i.e.
$$ {\bf T}^e = - \left( {\bf \nabla d} \right) \cdot {\partial W^d \over 
\partial \left( {\bf d \nabla} \right)} \eqno (10) $$

It is suggested in the Oseen-Frank-Nehring-Saupe theory that
$$ \eqalignno{ W^d =&
k_1 \left( {\bf\nabla} \cdot {\bf d} \right) - k_2 \left( {\bf d} \cdot
{\bf\nabla} \times {\bf d} \right) + {1\over 2} 
\left( k_{11}-2k_{13} \right)
\left( {\bf\nabla} \cdot {\bf d} \right)^2 + & \cr
& + {1\over 2} k_{22} \left( {\bf d} \cdot {\bf\nabla} \times {\bf d}
\right)^2 + {1\over 2} \left( k_{33} + 2k_{13} \right) \left( {\bf d}
\cdot {\bf\nabla d} \right)^2 - & \cr
& - \left( k_{12} - k_{23} \right) \left( {\bf d} \cdot {\bf\nabla}
\times {\bf d} \right) \left( {\bf\nabla} \cdot {\bf d} \right) + k_{13}
{\bf\nabla} \cdot \left( {\bf d\nabla} \cdot {\bf d} \right) - & \cr  
& - {1\over 2} \left( k_{22} + k_{24} \right) {\bf\nabla} \cdot \left(
{\bf d\nabla} \cdot {\bf d} - {\bf d} \cdot {\bf\nabla d} \right) - & \cr
& - k_{23} {\bf\nabla} \cdot \left[ \left( {\bf d} \cdot {\bf\nabla} 
\times {\bf d} \right) {\bf d} \right] & (11) \cr} $$
where $ k_i $'s are the elastic moduli 
(Frank's notation [5] is used in this paper), 
among which the relationship 
$ k_{24} = {1 \over 2} \left( k_{11} - k_{22} \right) $ holds.

In relation (11), the first term originally represented the splay structure 
resulting from the change of symmetry of the material, 
although liquid crystals having 
ferroelectric features are rarely found in nature. The elastic 
moduli $ k_2 $, $ k_{12} $ and $ k_{23} $ are relevant to the chirality of 
the molecules. The three squared terms stand for, in sequence, the splay, 
twist and bending modes of the elastic deformation. The last three terms 
relate to boundary distortions.

The viscous stress is supposed to be a linear function of the rate of 
strain tensor $ \bf S $ and the rotational velocity of the director 
$ \bf N $ [9,10]
$$ {\bf T}^v = \alpha_1 \left( {\bf d} \cdot {\bf S} 
\cdot {\bf d} \right) {\bf dd} + \alpha_2 {\bf dN} + 
\alpha_3 {\bf Nd} + \alpha_4 {\bf S} + \alpha_ 5 {\bf d} 
\left( {\bf S} \cdot {\bf d} \right) + \alpha_6 
\left( {\bf S} \cdot {\bf d} \right) {\bf d} \eqno (12) $$
in which the independent scalar variables such as the mass density 
and the temperature have been absorbed in the viscosity coefficients. 
The viscosity coefficients satisfy the relationship [11]
$$ \alpha_2 + \alpha_3 = \alpha_5 - \alpha_6 \eqno (13) $$

The two considered variables are given by
$$ {\bf S} = {1\over 2} \left( {\bf v \nabla} + {\bf \nabla v} \right) 
\eqno (14) $$
and
$$ {\bf N} = {\bf\dot d} - {1\over 2} {\bf\nabla} \times \left( {\bf v} - 
{\bf\dot r} \right) \times {\bf d} \eqno (15) $$
where $ \bf r $ is the position vector, the overdot indicates the 
convected derivative.

The movement of the director is governed by the equation [9,10]
$$ {d \over dt} 
\int\!\!\!\int\limits_R\!\!\!\int \tilde\rho {\bf\dot d} \, dR = 
\int\!\!\!\int\limits_R\!\!\!\int \left( {\bf G} + {\bf g} \right) \, dR + 
\int\limits_s\!\!\!\int d {\bf s} \cdot {\bf\Pi} \eqno (16) $$
where $ \tilde\rho $ is the moment of inertia per unit volume 
($ \rm ML^{-1} $), $ \bf G $ and $ \bf g $ are respectively the 
external director body force and the intrinsic director body force, 
having the dimensions of torque per unit volume ($ \rm ML^{-1}T^{-2} $), 
and $ \bf\Pi $ is the intrinsic stress (with the dimensions of 
$ \rm ML^{-2} T^{-2} $).

The magnetic field and electric field forces may, respectively, 
be obtained from the magnetization energy [9] 
$$  W^m = -{1\over 2} {\bf M} \cdot {\bf H} \eqno (17) $$
and the polarization energy [29]
$$  W^p = -{1\over 2} {\bf P} \cdot {\bf E} \eqno (18) $$
where $ \bf P $ is the electric dipole moment.

The constitutive relationship of $ \bf P $ linear with 
the electric field is given as [9] 
$$ {\bf P} = {\epsilon_\perp - 1 \over 4\pi} {\bf E} 
+ {\epsilon_\parallel - \epsilon_\perp \over 4\pi} 
\left( {\bf d} \cdot {\bf E} \right) {\bf d} = {\epsilon_\perp - 
1 \over 4\pi} {\bf E} + {\epsilon_a \over 4\pi} \left( {\bf d} \cdot {\bf E} 
\right) {\bf d} \eqno (19) $$
where $ \epsilon_\parallel $ and $ \epsilon_\perp $ are the constant
dielectric permeabilities with the interpretations indicated, and 
$ \epsilon_a $ is the dielectric anisotropy.

With the help of relations (6) and (17)-(19), 
the field-generated forces defined by [9] 
$$ {\bf G}^{m,\,e} = {\partial W^{m,\,p} \over \partial {\bf d}} 
\eqno (20) $$
are interpreted as
$$ {\bf G}^m = -\chi_a \left( {\bf d} \cdot {\bf H} \right) {\bf H} 
\eqno (21) $$
$$ {\bf G}^e = -{\epsilon_a \over 4\pi} \left( {\bf d} \cdot {\bf E} \right) 
{\bf E} \eqno (22) $$

The intrinsic body force $ \bf g $ consists also of three parts 
$$ {\bf g} = \lambda {\bf d} + {\bf g}^e + {\bf g}^v \eqno (23) $$
where $ \lambda $ is a constant.

The conservative and non-conservative body forces exerted on the 
director are [9] respectively 
$$ {\bf g}^e = -{\partial W^d \over \partial {\bf d}} \eqno (24) $$
and
$$ {\bf g}^v = \gamma_ 1 {\bf N} + \gamma_2 {\bf S} \cdot {\bf d} 
\eqno (25) $$
where $ \gamma_1 $ and $ \gamma_2 $ are the viscosity 
coefficients relevant to the intrinsic motion. 

Relationships between the $ \gamma $ coefficients in Eq.(25) and 
the $ \alpha $ coefficients in Eq.(12) are [9] 
$$ \gamma_1 = \alpha_3 - \alpha_2, \qquad  
 \gamma_2 = \alpha_6 - \alpha_5 \eqno (26) $$ 

The stress $ \bf\Pi $ has only a conservative part [9]
$$ {\bf\Pi} = {\partial W^d \over \partial \left( {\bf d \nabla} \right)} 
\eqno (27) $$

The extra-body force $ {\bf g}^e + {\bf g}^v $ and the extra-stress 
$ \bf\Pi $ generate a torque on the director.

\medskip
\leftline{\sl 2.3. Energy balance}
\smallskip

The general form of the energy conservation law in the continuum is [9]
$$ \eqalignno{ & {d\over dt} \int\!\!\!\int\limits_R\!\!\!\int {\rho\over 2} 
{\bf v \cdot v} + {\tilde\rho \over 2} {\bf\dot d} \cdot {\bf\dot d} 
+ \rho U \, dR &  \cr
& \hskip 1cm = \int\!\!\!\int\limits_R\!\!\!\int \left( {\bf F} \cdot 
{\bf v} + {\bf G} \cdot {\bf\dot d} + Q^h \right) \, dR+ & \cr
& \hskip 1.5cm + \int\limits_s\!\!\!\int d {\bf s} \cdot \left( {\bf T} \cdot 
{\bf v} + {\bf \Pi} \cdot {\bf\dot d} - {\bf J}^h \right) & (28) \cr} $$
where $ U $ is the internal energy per unit mass, $ Q^h $ is the heat supply 
per unit volume per unit time, and $ {\bf J}^h $ is the heat current 
flowing out of the volume per unit area per unit time.

For uniaxial liquid crystals, the heat flux is supposed to have a linear 
relationship with the thermal field [9]
$$ {\bf J}^h = \beta_0 {\bf\nabla} \Theta + \beta_1 \left( {\bf d} \cdot 
{\bf\nabla} \Theta \right) {\bf d} \eqno (29) $$
where $ \Theta $ is the temperature and $ \beta_0 $ and $ \beta_1 $ 
are the heat conductivities. 

\bigskip
\leftline{\bf 3. Adaptation to membranes}

Now we will adapt the Ericksen-Leslie dynamic theory to a bilayer membrane. 
Unless otherwise stated, the Latin suffix runs over 1, 2 and 3, 
whilst that the Greek suffixes run over 1 and 2. Italic symbols
indicate the components of 
a vector or tensor in the moving local frame (see below) and block symbols 
represent those in fixed global coordinates (inertial reference system). 
Summation convention is assumed throughout.

As suggested in the theory of thin shells, one defines a geometric 
surface situated at the middle position in the membrane and names it the 
``middle surface" (see Fig 1). The word ``surface" in the following means, 
uniquely, the middle surface. 

Let $ \bf r $ be the vector from an origin in Euclidean space to a point 
in the middle surface. Any spatial position in the membrane is then 
determined by
$$ {\rm x^i} = {\rm r^i} \left( \theta^\alpha , t \right) + \theta^3
{\rm n^i} \left( \theta^\alpha , t \right) \eqno (30) $$
where $ {\rm x^i} $ denotes the spatial coordinates, $ \theta^i $ are 
arbitrary parameters and $ {\rm n^i} $ is the normal vector to the middle 
surface.

If $ \theta^1 $ and $ \theta^2 $ are identified with the curvilinear
coordinates on the middle surface, then the local base vectors are given by 
$$ {\bf e}_\alpha = {\partial {\bf r} \over \partial \theta^\alpha} 
\qquad {\bf e}_3 = {\bf n} = { {\bf e}_1 \times {\bf e}_2 \over 
\left\vert {\bf e}_1 \times {\bf e}_2 \right\vert} = 
{1 \over \sqrt{ a}} {\partial {\bf r} \over \partial \theta^1} 
\times { \partial {\bf r} \over \partial \theta^2} \eqno (31) $$
where $ a $ is the determinant of the metric tensor
$$ a_{\alpha\beta} = {\bf e}_\alpha \cdot {\bf e}_\beta \eqno (32) $$ 

The second- and the third-order magnitude of the surface are
$$ b_{\alpha\beta} = - { \partial {\bf e}_3 \over \partial \theta^\alpha} 
\cdot {\bf e}_\beta = {\partial {\bf e}_\alpha \over \partial 
\theta^\beta} \cdot {\bf e}_3 \eqno (33) $$
$$ c_{\alpha\beta} = {\partial {\bf e}_3 \over \partial \theta^\alpha} 
\cdot {\partial {\bf e}_3 \over \partial \theta^\beta} \eqno (34) $$

For uniquely determined surface, the fundamental magnitudes satisfy 
the identity 
$$ c_{\alpha\beta} - 2Hb_{\alpha\beta} + K a_{\alpha\beta} = 0 \eqno (35) $$
where 
$$ 2H = a^{\beta\alpha} b_{\alpha\beta} \eqno (36) $$
is the first curvature (mean curvature) and 
$$ K = { \left\vert b_{\alpha \beta} \right\vert \over \left\vert a_{\alpha
\beta} \right\vert} = {b \over a} \eqno (37) $$
the second curvature (gaussian curvature).

The director vector $ \bf d $ (a unit vector parallel to the long molecular
axes) is considered to radiate from the middle surface on the side to which 
the normal vector is positive (Fig. 1).

The molecules displace in the Euclidean space at a velocity $ \bf v $. 
With respect to the local bases $ {\bf e}_i $ which move with the
surface, the velocity has the components 
$$ v^\alpha = \dot \theta^\alpha, \qquad v^3 = 0 \eqno (38) $$
because the particle flow is supposed to be parallel to the surface.
With respect to the spatial coordinates $ {\rm x^i} $ the velocity is given by
$$ {\rm v^i} = {\rm \dot x^i} = t^{\rm i}_\alpha v^\alpha + {\rm \dot r^i} 
\eqno (39) $$
where
$$ t^{\rm i}_\alpha ={\partial {\rm r^i} \over \partial \theta^\alpha} 
\eqno (40) $$
is a time-dependent hybrid tensor. 
The tensor $ t^{\rm i}_\alpha $ is used to associate the flexible and 
movable surface with the inertial reference system. For example, if the 
director is known by its components $ d^k $ with respect to the local bases, 
then its components in the global coordinates must be 
$ {\rm d^k} = t^{\rm k}_\alpha d^\alpha + {\rm n^k} d^3 $. 
Conversely, while one has the 
expansion of a field vector at a certain position of the middle surface with 
respect to the fixed coordinates, say, $ {\rm E_k} $, one knows 
naturally, with respect to the moving frame, its tangential parts $ E_\alpha 
= t^{\rm k}_\alpha {\rm E_k} $ and its normal part $ E_{(n)} = E_3 = 
{\rm E_k n^k} $. The subscript $ (n) $ denotes the normal part of the vectors.

One defines the varying configuration of the surface in a way that any 
point of the surface at the considered instant 
$ \left( \theta^1, \theta^2, t \right) $ is the endpoint of a translation 
of the point which occupied the same curvilinear coordinates on the surface 
at the preceding instant $ \left( \theta^1, \theta^2, t^\circ \right) $ along 
the normal vector $ {\bf n}^\circ $. The displacement speed of the surface 
is thus determined by
$$ {\rm \dot r^i} = \left( {\partial \rm r^{\rm i} \over \partial t} \right)
_{\theta^1, \theta^2} = w \rm n^{\rm i} \eqno (41) $$
where $ w $ is the rate of displacement.

Across the thickness of the membrane the non-uniformities, such as that of 
the mass, the temperature and the internal energy, as well as that of the 
velocity and the stress, are supposed negligible. (The assumption $ v^\alpha
_{\;,3}=0 $ implies that the two molecular layers do not mutually slip.) It 
follows that the differentiation of any tensor property of the material is 
a covariant differentiation along the surface. The surface derivative 
operator is defined by [30]
$$ {\bf\nabla}_s = {\bf e}^\beta {\partial \over \partial \theta^\beta} 
\eqno (42) $$

Replacing the three-dimensional operator $ \bf\nabla $ in Eq.(11) by the
two-dimensional operator $ {\bf\nabla}_s $ and multiplying both sides 
of the expression by the membrane thickness $ h $ yields the surface density 
of elastic energy of the membrane. To the first-order terms (to small 
tilting angle of the director to the normal), it is
$$ \eqalignno{ W^d_{(1)} & = -k'_1 \left( 2H - \partial_\sigma d^\sigma -
{1\over 2a} d^\sigma \partial_\sigma a \right)+ k'_2 \varepsilon^{\alpha
\beta} \partial_\beta d_\alpha + \cr
& \hskip 5mm + k'_{11} H \left( 2H - \partial_\sigma d^\sigma - {1\over 2a}
d^\sigma \partial_\sigma a \right) - \cr
& \hskip5mm - \left( k'_{22} + k'_{24} \right) \left[ K - 2H \left( 
\partial_\sigma d^\sigma + {1 \over 2a} d^\sigma \partial_\sigma a \right)
+ d_{\alpha, \, \beta(\tau)} b^{\beta\alpha} \right] - \cr
& \hskip5mm - 2k'_{12} H \varepsilon^{\alpha\beta} \partial_\beta d_\alpha -
2k'_{13} d^\beta \partial_\beta H & (43) \cr} $$
where the symbol $ \partial_\alpha $ denotes the partial derivative 
with respect to the parameter $ \theta^\alpha $, $ \varepsilon^{\alpha\beta} $ 
is the two-parametric permutation tensor, the prime indicates the surface 
covariant derivative and the subscript $ (\tau) $ the tangential part (see 
Appendix A). Here the elastic moduli in Eq.(11) have been replaced by
$$ k'_{i} = hk_{i}, \qquad k'_{ij} = hk_{ij} \eqno (44) $$

Relationship (43) includes the energy contributed by 
the non-uniform orientation 
of the molecules, which was emphasized to be important but excluded from 
consideration in the Helfrich's theory [7].

Letting $ {\bf d} = {\bf n} $, one recovers Helfrich's formula [7]
$$ W^d_{(0)} = k_c H + k'_c H^2 + \bar k_c K \eqno (45) $$
where
$$ k_c = - 2k'_1, \qquad k'_c = 2k'_{11}, \qquad 
\bar k_c= - \left( k'_{22} + k'_{24} \right) \eqno (46) $$

Hence the framework encompasses Helfrich's theory as a special case 
where the long molecular axes are parallel to the normal vector.
Recently, Ou-Yang and Liu deduced Eq.(45) from formula (11) [31].

The first term in Eq.(43) and in Eq.(45) now stands for 
the non-symmetry of the
two molecular leaves [7]. The last three terms in Eq.(43)
and the last term in Eq.(45) 
are actually meaningful only when the membrane has a free edge such as 
a helical strips (e.g. [32,33]). Otherwise, the orientation of 
the molecules at the 
periphery is determined by the veins of the solid boundary on which the 
membrane is braced rather than the internal molecular field. For a closed 
vesicle, these terms are vacuous of physical content.

The following kinematic theorem will frequently be used later, which
is, in fact, an analogue of the Reynolds' transport theorem 
in two-dimensional fluids [28]. 

If $ \phi $ is any function of position on a surface and of time, which
can be any scalar or tensor component, 
and $ s $ is a material part of the surface, then 
$$ {d \over dt} \int\limits_s\!\!\!\int \phi \, ds = \int\limits_s\!\!\!\int 
\left[ \dot\phi + \phi \left( v^\alpha_{\;,\,\alpha} + {\dot a \over 2a} \right) 
\right] \, ds \eqno (47) $$
where $ v^\alpha_{\;,\,\alpha} + \dot a / (2a) $ is the dilation of area.

The change in area of the membranes arises from, besides extension and 
compression, tilting of the molecules also [34-37]. For the surface 
configuration defined above, the rate of change of area is related to the 
displacement speed by [38]
$$ {\dot a \over 2a} = -2Hw \eqno (48) $$
If there is neither a source nor a sink in the membrane, then
$$ v^\alpha_{\;,\,\alpha} = 0 \eqno (49) $$

On account of (48) and (49), one rewrites (47) as
$$ {d \over dt} \int\limits_s\!\!\!\int \phi \, ds = \int\limits_s\!\!\!\int 
\left( \dot\phi - 2Hw\phi \right) \, ds \eqno (50) $$

\medskip
\leftline{\sl 3.1. Mass balance}
\smallskip

Let $ \gamma $ be the surface density of mass. The mass conservation 
principle requires
$$ {d \over dt} \int\limits_s\!\!\!\int \gamma \, ds = 0 \eqno (51) $$

Recalling theorem (50) and taking off the integration one obtains
$$ \partial_t \gamma + v^\beta \partial_\beta \gamma -2Hw\gamma = 0 
\eqno (52) $$
where the convected derivative $ d_t = \partial_t + v^\beta \partial_\beta $
has been taken into consideration.

If $ \xi_{(i)} $ is the mass percentage of the $ ith $ species inlaid in
the membrane, then the balance of this species gives
$$ \partial_t \xi_{(i)} + v^\alpha \partial_\alpha \xi_{(i)} - 2Hw \xi_{(i)}
= Q_{(i)} - J^\alpha_{(i),\,\alpha} \eqno (53) $$
where $ Q_{(i)} $ is the surface chemical source and $ {\bf J}_{(i)} $ the 
surface diffusion flux of the species.

\medskip
\leftline{\sl 3.2. Conservation of momentum}
\smallskip

The net force on an arbitrary area with periphery $ l $ resolved in the 
direction $ m_\beta $ is given by 
$$ {d \over dt} \int\limits_s\!\!\!\int \gamma w \, ds = 
\int\limits_s\!\!\!\int F^3 \, ds + \oint\limits_l m_\beta T^{\beta 3} \, dl 
\eqno (54) $$
$$ {d \over dt} \int\limits_s\!\!\!\int \gamma v^\alpha \, ds = 
\int\limits_s\!\!\!\int F^\alpha \, ds + \oint\limits_l m_\beta 
T^{\beta\alpha} \, dl \eqno (55) $$

Making use of theorem (50) and Eq.(52), one obtains Cauchy's equation 
for the displacement movement of the surface
$$ \gamma \dot w = F^3 + T^{\beta 3}_{\;,\,\beta} \eqno (56) $$ 
and that for the internal flow of the particles
$$ \gamma \dot v^\alpha = F^\alpha + T^{\beta\alpha}_{\;,\,\beta} 
\eqno (57) $$

In view of the perturbation analysis, one expands Eqs.(56) and (57) below 
on the assumption that the tilt angle of the director with respect to the 
normal vector is small. All the expressions in the previous section are 
valid once the spatial operator $ {\bf\nabla} $ is replaced everywhere by 
the surface operator $ {\bf\nabla}_s $ and the energy densities Eqs.(11), 
(17) and (18) are multiplied by the membrane thickness. The details of 
the derivation are given in Appendix C.

As is shown in Eq.(3), the body force $ \bf F $ consists of the gravitational 
attraction and electromagnetic field-generated forces. The expansion of 
these vectors expressed in Eqs.(4)-(7) with respect to the local bases gives
$$ \eqalignno{ {F^g}_3 & = \gamma \hat g_{(\rm n)} & (58) \cr 
{F^g}_\alpha & = \gamma {\rm\hat g}_{\rm i} t^{\rm i}_\alpha & (59) \cr 
{F^m}_3 & = \rm H_{\rm j,i} \rm n^{\rm j} \left[ \chi_\perp \left(
\rm H^{\rm i} - \rm H_{(\rm n)} \rm n^{\rm i} \right) + \chi_a 
\rm H_{(\rm n)} t^{\rm i}_\beta d^\beta \right] & (60) \cr 
{F^m}_\alpha & = \rm H_{\rm j,i} t^{\rm j}_\alpha \left[ \chi_\perp \left(
\rm H^{\rm i} - \rm H_{(\rm n)} \rm n^{\rm i} \right) + \chi_a 
\rm H_{(\rm n)} t^{\rm i}_\beta d^\beta \right] & (61) \cr 
{F^e}_3 & = \gamma_e E_{(n)} & (62) \cr
{F^e}_\alpha & = \gamma_e \rm E_{\rm i}t^{\rm i}_\alpha & (63) \cr } $$
where $ \gamma_e $ is the surface density of electric charges.

The stress tensor $ \bf T $ consists of (as shown in (8)) an 
isotropic part, a conservative part and a non-conservative part. The 
application of the surface divergence operator to the isotropic stress yields
$$ {T^i}^{\beta\alpha}_{\;,\,\beta} = - a^{\beta\alpha} \partial_\beta
\sigma \eqno (64) $$
where $ \sigma $ is identified with the surface tensor.
The elastic restoring force of the membrane arises from the tilt of the 
molecules and the curvature of the sheet. When the long molecular axes have 
a small angle to the normal vector while the shape of the membrane is flat, 
then the surface force reads
$$ \eqalignno{ {T^e}^{\beta 3}_{\;,\, \beta} & = 0 & (65) \cr 
{T^e}^{\beta\alpha}_{\;,\, \beta} & = - k'_1 a^{\beta\sigma} 
d^\alpha_{\;,\, \sigma\beta (\tau)} + k'_2 \varepsilon^{\alpha\sigma} 
a^{\beta\gamma} d_{\sigma ,\,\gamma\beta (\tau)} & (66) \cr} $$ 
Conversely, when the membrane is flexible while the long molecular axes 
remain parallel to the normal vector, then the surface force is given by 
$$ \eqalignno{ {T^e}^{\beta 3}_{\;,\, \beta} & = 
\left( k'_1 - 2H k'_{11} \right) \left( 4H^2 - 2K \right) + 
\left( k'_{22} + k'_{24} \right) 2HK & (67) \cr
{T^e}^{\beta\alpha}_{\;,\, \beta} & = 2k'_1 a^{\alpha\beta} \partial_\beta 
H - 2k'_2 \varepsilon^{\alpha\beta} \partial_\beta H - 2k'_{11}
\left( b^{\alpha\beta} + 2H a^{\alpha\beta} \right) \partial_\beta H + \cr
& \hskip5mm + \left( k'_{22} +k'_{24} \right) a^{\alpha\beta} \partial_\beta 
K +2k'_{12} \varepsilon^{\alpha\beta} \left( b^\sigma_\beta \partial_\sigma 
H + \partial_\beta H^2 \right) & (68) \cr} $$ 

One considers next the viscous stress.
It is easy to verify that for the two-dimensional fluid the rate of strain
given by Eq.(14) is 
$$ {\bf S} = {1\over 2} \left( v_{\alpha,\,\beta} + v_{\beta,\,\alpha} - 
2wb_{\alpha\beta} \right) {\bf e}^\alpha {\bf e}^\beta \eqno (69) $$
and the rotational velocity of the director given in Eq.(15) 
$$ {\bf N} = \left[ {1\over 2} v^\beta b^\alpha_\beta + \dot d^\alpha -
{1\over 2} \varepsilon^{\alpha\beta} d_\beta \left( \varepsilon^{\lambda\mu} 
\partial_\lambda v_\mu \right) \right] {\bf e}_\alpha + {1\over 2} v^\beta
b_{\beta\alpha} d^\alpha {\bf e}_3 \eqno (70) $$
where
$ \dot d^\alpha = \partial_t d^\alpha + v^\sigma 
d^\alpha_{\;,\,\sigma(\tau)}$ . 
In writing Eq.(69) one took into account the formula [38] 
$$ \dot a_{\lambda\mu} = - 2wb_{\lambda\mu} \eqno (71) $$
(see Appendix C.)
Substituting expressions (69) and (70) into (12) gives the viscous resistance 
in a two-dimensional fluid. In the limiting case $ 2H = K = 0 $ it is given 
by 
$$ \eqalignno { 
{T^v}^{\beta 3}_{\;,\,\beta} =
& \alpha_3 \biggl\{ \partial_t \left( \partial_\sigma d^\sigma + {1\over 2a} 
d^\sigma \partial_\sigma a \right) + v^\alpha_{\;,\,\beta} 
d^\beta_{\;,\,\alpha(\tau)} + v^\alpha d^\beta_{\;,\,\alpha\beta(\tau)} - 
\biggr. \cr
& \hskip 0.6cm \biggl. - {1\over 2} \varepsilon^{\beta\sigma} 
\left( d_\sigma \partial_\beta + \partial_\beta d_\sigma \right)
\left( \varepsilon^{\lambda\mu} \partial_\lambda v_\mu \right) \biggr\} + \cr
& + {\alpha''_6 \over 2} E^{\beta\alpha\lambda\mu} \left[ \left( 
v_{\mu,\,\lambda\beta} + v_{\lambda,\,\mu\beta} \right) d_\alpha + 
\left( v_{\mu ,\,\lambda} + v_{\lambda, \,\mu} \right) 
d_{\alpha,\,\beta (\tau)} \right] & (72) \cr
{T^v}^{\beta\alpha}_{\;,\,\beta} =
& {\eta\over 2} E^{\alpha\beta\lambda\mu} \left( v_{\lambda ,\,\mu\beta} +
v_{\mu ,\,\lambda\beta} \right) & (73) \cr} $$
and in the other limiting case $ {\bf d} \parallel {\bf n} $ it is given by 
$$ \eqalignno { 
{T^v}^{\beta 3}_{\;,\,\beta} = 
& - \left( \mu + \eta \right) 4H^2w + \eta \left[ 4Kw + \left( b^{\lambda\mu}
- Ha^{\lambda\mu} \right) \left( v_{\lambda,\,\mu} + v_{\mu,\,\lambda} 
\right) \right] + \cr
& + {\alpha_3 \over 2} \left[ v^\alpha_{\;,\,\beta} b^\beta_\alpha + 
v^\beta \partial_\beta \left( 2H \right) \right] & (74) \cr 
{T^v}^{\beta\alpha}_{\;,\,\beta} =
& - \mu a^{\alpha\beta} \partial_\beta \left( 2Hw \right) + 
{\eta\over 2} E^{\alpha\beta\lambda\mu} \left( v_{\lambda ,\,\mu\beta} +
v_{\mu ,\,\lambda\beta} \right) - \cr
& - 2 \eta \left[ \partial_\beta w 
\left( b^{\alpha\beta} - H a^{\alpha\beta} \right) + w a^{\alpha\beta} 
\partial_\beta H \right] + \cr
& + \alpha_2 H v^\beta b^\alpha_\beta + {\alpha_3 \over 2} \left( 2H 
v_\beta b^{\beta\alpha} - K v^\alpha \right) & (75) \cr} $$   
where $ \mu $ is the dilatation viscosity, $ \eta $ is 
the shear viscosity and 
$ E^{\alpha\beta\lambda\mu} $ is the fourth-order tensor 
$ E^{\alpha\beta\lambda\mu} = a^{\alpha\lambda} a^{\beta\mu} + a^{\alpha\mu}
a^{\beta\lambda} - a^{\alpha\beta} a^{\lambda\mu} $. 
In deducing (72)-(75), formula (48) was used.

The momentum balance for the intrinsic motion over the area $ s $ 
Eq.(16) is written as 
$$ \eqalignno{ 
0 & = \int\limits_s\!\!\!\int \left( G^3 + g^3 \right) \, ds
+ \oint\limits_l m_\beta \Pi^{\beta 3} \, dl & (76) \cr
{d\over dt} \int\limits_s\!\!\!\int \tilde\gamma \dot d^\alpha & =
\int\limits_s\!\!\!\int \left( G^\alpha +g^\alpha \right) \, ds + 
\oint\limits_l m_\beta \Pi^{\beta\alpha} \, dl & (77) \cr } $$ 
where $ \tilde\gamma = \gamma h $. In the usual way, Eqs.(76) and (77) 
lead directly to the point expression of the law
$$ \eqalignno{ 0 & = G^3 + g^3 + \Pi^{\beta 3}_{\;,\beta} & (78) \cr
\tilde\gamma \ddot d^\alpha & = G^\alpha + g^\alpha + \Pi^{\beta\alpha}_
{\;,\,\beta} & (79) \cr} $$ 

When a polar fluid is exposed to the external magnetic field, the director 
is reoriented by the force $ {\bf G}^m $, as given by Eq.(21) 
$$ \eqalignno{
{G^m}_3 & = - \chi_a H_{(n)} 
\left( {\rm H}_{\rm i} t^{\rm i}_\sigma d^\sigma + H_{(n)} \right) & (80) \cr 
{G^m}_\alpha & = - \chi_a \rm H_{\rm j} t^{\rm j}_\alpha
\left( {\rm H}_{\rm i} t^{\rm i}_\sigma d^\sigma + H_{(n)} \right) 
& (81) \cr} $$ 
In an external electric field, it is subjected to the force $ {\bf G}^e $, 
as given by Eq.(22) 
$$ \eqalignno{
{G^e}_3 & = - {\epsilon_a\over 4 \pi} E_{(n)} 
\left( {\rm E}_{\rm i} t^{\rm i}_\sigma d^\sigma + E_{(n)} \right) & (82) \cr
{G^e}_\alpha & = - {\epsilon_a\over 4 \pi} {\rm E}_{\rm j} t^{\rm j}_\alpha
\left( {\rm E}_{\rm i} t^{\rm i}_\sigma d^\sigma + E_{(n)} \right) 
& (83) \cr} $$ 

For a two-dimensional fluid, the extra-body forces given by Eqs.(24) and 
(25), in the limiting case $ 2H = K = 0 $, are given by 
$$ \eqalignno{ 
{g^e}^3 = & - k'_2 \varepsilon^{\alpha\beta} \partial_\beta d_\alpha 
& (84) \cr 
{g^e}^\alpha = & - k'_{13} a^{\alpha\beta} \partial_\beta \left( \partial_ 
\sigma d^\sigma + {1\over 2a} d^\sigma \partial_\sigma a \right) 
+ k'_{23} a^{\alpha
\beta} \partial_\beta \left( \varepsilon^{\lambda\mu} \partial_\lambda d_\mu 
\right) & (85) \cr
{g^v}^3 = & 0 & (86) \cr
{g^v}^\alpha = & \gamma_1 \left( \dot d^\alpha + {1\over 2} 
\varepsilon^{\lambda\beta} \partial_\lambda v_\mu d_\beta \varepsilon
^{\beta\alpha} \right)+ {\gamma''_2\over 2} E^{\alpha\beta\lambda\mu} 
\left( v_{\lambda,\,\mu} + v_{\mu,\,\lambda} \right) d_\beta & (87) \cr} $$
and, in the other limiting case $ {\bf d} \parallel {\bf n} $,
$$ \eqalignno{
{g^e}^3 = & 0 & (88) \cr
{g^e}^\alpha = & k'_{13} a^{\alpha\beta} \partial_\beta \left( 2H \right) 
& (89) \cr
{g^v}^3 = & 0 & (90) \cr 
{g^v}^\alpha = & {\gamma_1\over 2} v^\beta b^\alpha_\beta & (91) \cr} $$
where $ \gamma''_2 $ is referred to the shear motion.

The surface force to which the director is subjected during the 
elastic deformation, given by Eq.(27), takes the form
$$ \eqalignno{ \Pi^{\beta 3}_{\;,\,\beta} = 
& k'_2 \varepsilon^{\beta\alpha} \partial_\beta d_\alpha & (92) \cr
\Pi^{\beta\alpha}_{\;,\,\beta} = 
& {1\over 2} \left( k'_{11} -k'_{22} \right) a^{\alpha\beta} \partial_\beta
\left( \partial_\sigma d^\sigma + {1\over 2a} d^\sigma \partial_\sigma a
\right) + \cr
& + k'_{22} \varepsilon^{\alpha\beta} \partial_\beta \left( \varepsilon
^{\lambda\mu} \partial_\lambda d_\mu \right) + \left( k'_{22} + k'_{24} 
\right) a^{\beta\gamma} d^\alpha_{,\,\gamma\beta(\tau)} - \cr
& - k'_{12} \left[ \varepsilon^{\alpha\beta} \partial_\beta \left(
\partial_\sigma d^\sigma + {1\over 2a} d^\sigma \partial_\sigma a \right) +
a^{\alpha\beta} \partial_\beta \left( \varepsilon^{\lambda\mu} \partial_
\lambda d_\mu \right) \right] & (93) \cr} $$
for $ 2H = K =0 $ and
$$ \eqalignno{ \Pi^{\beta 3}_{\;,\,\beta} =
& \left( k'_1 - 2Hk'_{11} \right) 2H + \left( k'_{22} + k'_{24} \right) 2K 
& (94) \cr
\Pi^{\beta\alpha}_{\;,\,\beta} = 
& - k'_{11} a^{\alpha\beta} \partial_\beta \left( 2H \right) + k'_{12} 
\varepsilon^{\alpha\beta} \partial_\beta \left( 2H \right) & (95) \cr} $$ 
for $ {\bf d} \parallel {\bf n} $. 

\medskip
\leftline{\sl 3.3. Energy balance}
\smallskip

The energy balance Eq.(28) for any material area of the membrane 
reads
$$ \eqalignno{ 
& {d\over dt} \int\limits_s\!\!\!\int {\gamma\over 2} \left(
v^\alpha v_\alpha + w^2 \right) + {\tilde\gamma\over 2} \dot 
d^\alpha \dot d_\alpha + \gamma U \, ds \cr
& \hskip 1cm = \int\limits_s\!\!\!\int \left( F^\alpha v_\alpha + F^3 w + 
G^\alpha \dot d_\alpha + Q^h \right) \, ds + \cr
& \hskip 1.4cm + \oint\limits_l m_\beta \left( T^{\beta\alpha} v_\alpha 
+ T^{\beta 3} w + \Pi^{\beta\alpha} \dot d_\alpha - {J^h}^\beta \right) \, dl 
& (96) \cr} $$
where $ Q^h $ is the rate of local production of heat per unit area and 
$ {\bf J}^h $ is the surface diffusion current across 
a unit linear element of 
the periphery of the considered area.

Taking account of Eqs.(56)-(57), and (78)-(79), one obtains from 
Eq.(96) a differential expression of the law 
$$ \gamma \dot U = Q^h - {J^h}^\alpha_{\;,\,\alpha} + T^{\beta\alpha} v_
{\alpha,\,\beta} + T^{\beta 3} \partial_\beta w + \Pi^{\beta\alpha} \dot d_
{\alpha,\,\beta} - g^\alpha \dot d_\alpha \eqno (97) $$

For practical uses, one replaces the internal energy by a testable 
physical quantity and obtains
$$ \eqalignno{ & \gamma C_l \dot\Theta - {T^e}^{\beta\alpha} v_{\alpha,\,
\beta} - {T^e}^{\beta 3} \partial_\beta w - \Pi^{\beta\alpha} {\dot d}
_{\alpha,\,\beta} + {g^e}^\alpha \dot d_\alpha \cr 
& \hskip 0.5cm = Q^h - {J^h}^\alpha_{\;,\,\alpha} + \left( - \sigma 
a^{\beta\alpha} + {T^v}^{\beta\alpha} \right) v_{\alpha,\,\beta} + 
{T^v}^{\beta 3} \partial_\beta w - {g^v}^\alpha \dot d_\alpha 
& (98) \cr} $$
where $ C_l $ is the heat capacity per unit mass in the static state 
[39], the convected derivative of the temperature is considered as 
$$ \dot\Theta = \partial_t \Theta + v^\alpha t^{\rm j}_\alpha 
\partial_{\rm j} \Theta + w {\rm n}^{\rm j} \partial_{\rm j} \Theta 
\eqno (99) $$
(see Appendix D.)

The non-vanishing terms of Eq.(98) containing $ {\bf T}^{e,\,v} $, 
$ {\bf g}^{e,\,v} $ and $ \bf \Pi $ are
$$ \eqalignno{ {T^e}^{\beta\alpha} v_{\alpha,\,\beta} = 
& - \left( k'_1 a^{\beta\sigma} d^\alpha_{\;,\,\sigma (\tau)} - k'_2 
a^{\beta\lambda} \varepsilon^{\alpha\mu} d_{\mu,\,\lambda}\right)
v_{\alpha,\,\beta} & (100) \cr
\Pi^{\beta\alpha} \dot d_{\alpha,\,\beta} = 
& \left( k'_1 a^{\beta\alpha} + k'_2 \varepsilon^{\beta\alpha} \right)
\dot d_{\alpha,\,\beta} & (101) \cr
{T^v}^{\beta\alpha} v_{\alpha,\,\beta} = 
& {\eta \over 2} E^{\alpha\beta\lambda\mu} \left( v_{\lambda,\,\mu} + 
v_{\mu,\,\lambda} \right) v_{\alpha,\,\beta} & (102) \cr
{T^v}^{\beta 3} \partial_\beta w = 
& \biggl\{ \alpha_3 \left[ \dot d^\beta + {1\over 2} \left( \varepsilon^
{\lambda\mu} \partial_\lambda v_\mu \right) \varepsilon^{\beta\alpha} 
d_\alpha \right] + \biggr. \cr
& \hskip 0.4cm \biggl. + {\alpha''_6 \over 2} E^{\beta\alpha\lambda\mu} 
\left( v_{\lambda,\,\mu} + v_{\mu,\,\lambda} \right) d_\alpha \biggl\}
\partial_\beta w & (103) \cr} $$
for $ 2H = K = 0 $, and
$$ \eqalignno{ {T^e}^{\beta\alpha} v_{\alpha,\,\beta} =
& \Bigl[ \left( k'_1 - 2Hk'_{11} \right) b^{\beta\alpha} + 
\left( k'_2 - k'_{12} \right) b^\beta_\gamma \varepsilon^{\gamma\alpha} + 
\Bigr. \cr
& \Bigl. \hskip 0.5cm + \left( k'_{22} + k'_{24} \right) K a^{\beta
\alpha} \Bigr] v_{\alpha,\,\beta} & (104) \cr
{g^e}^\alpha \dot d_\alpha = 
& k'_{13} \dot d_\alpha a^{\alpha\beta} \partial_\beta \left( 2H \right) 
& (105) \cr
\Pi^{\beta\alpha} \dot d_{\alpha,\,\beta} =
& \Bigl[ \left( k'_1 - 2Hk'_{11} \right) a^{\beta\alpha} + \left( k'_2 -
2Hk'_{12} \right) \varepsilon^{\beta\alpha} + \Bigr. \cr
& \hskip 2mm \Bigl. + \left( k'_{22} + k'_{24} \right) \left( 2H 
a^{\beta\alpha} - b^{\beta\alpha} \right) \Bigr] \dot d_{\alpha,\,\beta} 
& (106) \cr
{T^v}^{\beta\alpha} v_{\alpha,\,\beta} =
& \biggl\{ \left( \eta - \mu \right) 2Hw a^{\beta\alpha} + \biggr. \cr
& \hskip 0.5cm \biggl. + \eta \left[ {1\over 2} 
E^{\alpha\beta\lambda\mu} \left( v_{\lambda,\,\mu} + v_{\mu,\,\lambda} 
\right) - 2w b^{\beta\alpha} \right] \biggr\} v_{\alpha,\,\beta} & (107) \cr
{T^v}^{\beta 3} \partial_\beta w =
& - {\alpha_3 \over 2} v^\sigma b^\beta_\sigma \partial_\beta w & (108) \cr
{g^v}^\alpha \dot d_\alpha = 
& {\gamma_1 \over 2} v_\beta b^{\beta\alpha} \dot d_\alpha & (109) \cr} $$
for $ {\bf d} \parallel {\bf n} $. Here the covariant derivative of
$ \dot d_\alpha $ is given by 
$$ \dot d_{\alpha,\,\beta} = \partial_t \left( d_{\alpha,\,\beta(\tau)} 
\right) + v^\gamma_\beta d_{\alpha,\,\gamma(\tau)} + v^\gamma 
d_{\alpha,\,\gamma\beta(\tau)} \eqno (110) $$
(see Appendix C.)

For the local production of heat it is commonly written as [39] 
$$ Q^h = \sum_r \sum_k H_k \nu_{kr} b_r \eqno (111) $$
where $ H_k $ is the enthalpy of the $ kth $ species per unit mass, 
$ \nu_{kr} $ is the stoichiometric number of species $ k $ 
in chemical reaction $ r $ 
and $ b_r $ is the reaction rate of chemical reaction $ r $ referred to 
a unit area.

The divergence of the heat flux Eq.(29) is given 
for the two-dimensional fluid, by 
$$ \eqalignno{ {J^h}^\alpha_{\;,\,\alpha} = 
& - \kappa_\perp \left( \partial_{\rm j} \Theta \right)_{,\,\rm k} \left( 
{\rm g}^{\rm jk} - {\rm n}^{\rm j} {\rm n}^{\rm k} \right) + \kappa_a 2H 
\left( {\rm n}^{\rm j} \partial_{\rm j} \Theta \right) + \cr
& + \kappa_a \biggl[ \left( t^{\rm j}_\beta \partial_{\rm j} 
\Theta \right) \left( 2H d^\beta + b^\beta_\alpha d^\alpha \right) -
{\rm n}^{\rm j} \left( \partial_{\rm j} \Theta \right)_{,\,\rm k} t^{\rm k}
_\alpha d^\alpha - \biggr. \cr
& \hskip 1.1cm \biggl.- \left( {\rm n}^{\rm j} \partial_{\rm j} \Theta \right) 
\left( \partial_\sigma d^\sigma + {1\over 2a} d^\sigma \partial_\sigma a
\right) \biggr] & (112) \cr} $$
where $ \kappa_\perp $ is the heat conductivity in the direction 
perpendicular to the director vector and $ \kappa_a \equiv \kappa_\parallel 
- \kappa_\perp $ the thermal anisotropy.

In the above derivations, one takes an arbitrary three-dimensional 
coordinate frame as the inertial reference system and two arbitrary lines on
the middle surface as the curvilinear coordinates. For most purposes, however,
one may choose a simple coordinate system. For instance, if one defines the 
two principal curvature lines as the curvilinear coordinates, which leads to
an orthogonal conjugate system, then the expressions will be much simplified.

\bigskip
\leftline{\bf 4. Concluding remarks}
\medskip

By adapting the electrohydrodynamic theory of uniaxial liquid crystals to 
thin films, one obtains a group of dynamic equations, 
including basic Eqs.(52)-(53), (56)-(57), (78)-(79) and (98), as well
as relevant expressions (58)-(70), (72)-(75), (80)-(95), 
(100)-(109) and (111)-(112), which may conveniently be used in the study 
of hydrodynamic and electrodynamic phenomena of biomembrane matrix. The 
surface density of 
elastic energy Eq.(43) may be used to the hydrostatic studies of 
membranes with particular regard to molecule tilting.

\bigskip
\leftline{\bf Acknowledgements}
\medskip

This work is supported by the National Foundation for Natural Sciences and
by the Foundation of the Ministry of Education of China.

The author is grateful to Dr. Y. K. Lau for reading the manuscript and 
suggesting many changes in the presentation to make it more readable.

\bigskip

\leftline{\bf Appendix A. Preliminaries}  

\medskip

The Gauss formula is given by 
$$ \eqalignno{ {\partial {\bf e}_\alpha\over \partial \theta^\beta} & =
\left\{ \matrix{ \mu \cr \alpha \; \beta \cr} \right\} {\bf e}_\mu +
b_{\alpha\beta} {\bf e}_3 \cr
{\partial {\bf e}^\alpha\over \partial \theta^\beta} & =
- \left\{ \matrix{ \alpha \cr \mu \; \beta \cr} \right\} {\bf e}^\mu +
b^\alpha_\beta {\bf e}^3 & (A1) \cr} $$
and the derivative of the normal vector
$$ {\partial {\bf e}_3 \over \partial \theta^\beta} =
- b_{\beta}^\mu {\bf e}_\mu = - b_{\beta\mu} {\bf e}^\mu \eqno (A2) $$
where $\left\{ \matrix{\alpha \cr \sigma \; \beta \cr} \right\} $ is the 
second kind of Christoffel symbol.

Some surface differential invariants of the director vector are given as
$$ \eqalignno{ {\bf \nabla}_s {\bf d} 
& = \left( d^\alpha_{\;,\,\beta(\tau)} - d^3 b_\beta^\alpha \right) {\bf e}^
\beta {\bf e}_\alpha + \left( \partial_\beta d^3 + d^\sigma b_{\sigma\beta} 
\right) {\bf e}^\beta {\bf e}_3 & \cr
& = d^\alpha_{\;,\,\beta} {\bf e}^\beta {\bf e}_\alpha + d^3 _{\;,\,\beta} 
{\bf e}^\beta {\bf e}_3 & (A3) \cr
{\bf\nabla}_s \cdot {\bf d} & = d^\alpha_{\;,\,\alpha(\tau)} - 2H d^3 = 
d^\alpha_{\;,\,\alpha} & (A4) \cr
{\bf\nabla}_s \times {\bf d} & = \varepsilon^{\alpha\beta} \left[
\left( \partial_\beta d_3 + d^\sigma b_{\sigma\beta} \right) {\bf e}_\alpha
- \partial_\beta d_\alpha {\bf e}_3 \right] & \cr
& = \varepsilon^{\alpha\beta} \left(d_{3,\,\beta} {\bf e}_\alpha -\partial_
\beta d_\alpha {\bf e}_3 \right) & (A5) \cr} $$
where $ \varepsilon^{\alpha\beta} $ is a two parametric permutation
tensor, having the components $ \varepsilon^{11} = \varepsilon^{22} = 0 $,
$ \varepsilon^{12} = - \varepsilon^{21} = 1 / \sqrt a $. The subscript 
$ (\tau) $ distinguishes the tangential part 
$$ d^\alpha_{\;,\,\beta(\tau)} = {\partial_\beta d^\alpha} + d^\sigma
\left\{\matrix{\alpha \cr \sigma\;\beta \cr} \right\} \eqno (A6) $$
from the whole covariant derivative
$$ \eqalignno{ d^\alpha_{\;,\,\beta} & = \partial_\beta d^\alpha + d^\sigma
\left\{ \matrix{ \alpha \cr \sigma \; \beta \cr} \right\} - d^3 
b^\alpha_\beta = d^\alpha_{\;,\,\beta (\tau)} - d^3 b^\alpha_\beta \cr
d^3_{\;,\,\beta} & = \partial_\beta d^3 + d^\sigma b_{\sigma\beta} 
& (A7) \cr} $$

Supposing $ \bf A $ is a field vector. The surface derivative of the vector 
is associated with the spatial derivative of the vector by   
$$ {\bf\nabla}_s {\bf A} = {\rm A}_{\rm k,\,j} t^{\rm j}_\alpha {\bf e}
^\alpha \left( t^{\rm k}_\beta {\bf e}^\beta + {\rm n}^{\rm k} {\bf e}^3 
\right) \eqno (A8) $$

Let $ \bf B $ be a second-order tensor ($ {\bf B} = {\bf T} $ or 
$ {\bf \Pi} $). The contraction of the third-order tensor 
$ {\bf \nabla}_s {\bf B} $ gives
$$ \eqalignno{ {\bf \nabla}_s \cdot {\bf B} = 
& \left( B^{\beta\alpha}_{\;,\,\beta(\tau)} - B^{\beta 3} b^\alpha_\beta -
2H B^{3 \alpha} \right) {\bf e}_\alpha + \cr
& + \left( B^{\beta 3}_{\;,\,\beta(\tau)} + B^{\beta\alpha}
b_{\alpha\beta} - 2H B^{33} \right) {\bf e}_3 & (A9) \cr} $$
with
$$ \eqalignno{ B^{\beta\alpha}_{\;,\,\beta(\tau)} 
& = \partial_\beta B^{\beta\alpha} + B^{\beta\alpha} 
{1 \over 2 a} \partial_\beta a 
+ B^{\beta\gamma} \left\{ \matrix{ \alpha \cr 
\gamma\;\beta \cr} \right\} \cr
B^{\beta 3}_{\;,\,\beta(\tau)}
& = \partial_\beta B^{\beta 3} + B^{\beta 3} 
{1\over 2a} \partial_\beta a & (A10) \cr} $$ 
The normal vector to the middle surface has the spatial components
$$ {\bf n}_{\rm i} = {1\over 2} \varepsilon_{\rm ijk} t^{\rm j}_\alpha t^
{\rm k}_\beta \varepsilon^{\alpha\beta} \eqno (A11) $$
where $ \varepsilon_{\rm ijk} $ is the three-parametric permutation tensor,
which is equal to $ \sqrt{\rm g} $ when $ {\rm ijk} $ 
is an even permutation of 
1, 2, 3, equal to $ - \sqrt{\rm g} $ when it is an odd permutation and $ 0 $ 
in other cases, $ \rm g $ being the determinant of the metric tensor of the 
spatial coordinates $ {\rm g}_{\rm ij} $.

\bigskip
\leftline{\bf Appendix B. Surface density of elasticity energy}
\medskip

Volume density of elasticity energy Eq.(11) can otherwise be expressed as
$$ \eqalignno{ W^d = & k_1 \left( {\bf\nabla} \cdot {\bf d}\right) - k_2 
\left( {\bf d} \cdot {\bf\nabla} \times {\bf d} \right) + {1\over 2} k_{11} 
\left( {\bf\nabla} \cdot {\bf d} \right)^2 + \cr 
& + {1\over 2} k_{22} \left( {\bf d} \cdot {\bf\nabla} \times {\bf d} 
\right)^2 + {1\over 2} \left( k_{33} + 2k_{13} \right) \left( {\bf d} \cdot
{\bf\nabla d} \right)^2 - \cr
& - {1\over 2} \left( k_{22} + k_{24} \right) \left[ \left( {\bf\nabla} \cdot 
{\bf d} \right)^2 - {\bf d \nabla} \cdot\cdot {\bf d \nabla} \right] - \cr 
& - k_{12} \left( {\bf\nabla} \cdot {\bf d} \right) \left( {\bf d} \cdot
{\bf\nabla} \times {\bf d} \right) + \cr
& + k_{13} {\bf d} \cdot {\bf\nabla} \left( {\bf\nabla} \cdot {\bf d} \right) 
- k_{23} {\bf d} \cdot {\bf\nabla} \left( {\bf d} \cdot {\bf\nabla} \times 
{\bf d} \right) & (A12) \cr} $$

Replacing $ \bf\nabla $ by the surface operator $ {\bf\nabla}_s $ 
defined in Eq.(42), and multiplying the moduli by the membrane thickness $ h $, 
one obtains from Eq.(A12) the surface density of the elasticity energy
$$ \eqalignno{ W^d = & - k'_1 \left( 2H - d^\alpha_{\;,\,\alpha (\tau)} 
\right) + k'_2 \varepsilon^{\alpha\beta} \partial_\beta d_\alpha + k'_{11} 
H \left( 2H - d^\alpha_{\;,\,\alpha(\tau)} \right) - \cr
& - \left( k'_{22} + k'_{24} \right) \left( K - 2Hd^\alpha_{\;,\,\alpha
(\tau)} + d_{\alpha,\,\beta (\tau)} b^{\beta\alpha} \right) - \cr
& - k'_{12} 2H \varepsilon^{\alpha\beta} \partial_\beta d_\alpha -
k'_{13} d^\beta \partial_\beta \left( 2H \right) + O \left( {d^\alpha}^2
\right) & (A13) \cr} $$
where $ k'_{i} = hk_{i}, \; k'_{ij} = hk_{ij} $.

\bigskip
\leftline{\bf Appendix C. Forces sustained in membranes}
\medskip
\leftline{\sl C1. Forces being exerted on the director}
\smallskip
\leftline{\sl C1.1. External forces}
For a two-dimensional fluid, the force to which 
the director is subjected to in an 
external magnetic field is given by 
$$ \eqalignno{ {\bf G}^m & = - \chi_a \left( \bf d \cdot \bf H \right) 
\bf H \cr
& = - \chi_a \left( {\rm H}_{\rm k} t_\sigma^{\rm k} d^\sigma + H_{(n)} 
\right) \left( {\rm H}_{\rm j} t_\alpha^{\rm j} {\bf e}^\alpha + 
H_{(n)} {\bf e}^3 \right) & (A14) \cr} $$
and that in an external electric field is given by 
$$ \eqalignno{ {\bf G}^e & = - {\epsilon_a\over 4\pi} \left({\bf d} \cdot 
{\bf E} \right) {\bf E} \cr 
& = - {\epsilon_a\over 4\pi} \left(  {\rm E}_{\rm k}
t_\sigma^{\rm k} d^\sigma + E_{(n)} \right) \left( {\rm E}_{\rm j}
t_\alpha^{\rm j} {\bf e}_\alpha + E_{(n)} {\bf e}^3 \right) & (A15) \cr} $$

\smallskip
\leftline{\sl C1.2. Extra body forces}
\smallskip
\noindent{{\sl C1.2.1 Elastic restoring force.}
Inserting Eq.(A12) in relation (24), one obtains}
$$ \eqalignno{ {\bf g}^e & = - {\partial W^d\over \partial {\bf d}} \cr
& = k_2 {\bf\nabla} \times {\bf d} - k_{22} \left( {\bf d} \cdot {\bf\nabla} 
\times {\bf d} \right) \left( {\bf\nabla} \times {\bf d} \right) - \cr
& \hskip 0.4cm - \left( k_{33} + 2k_{13} \right) {\bf d} \cdot \left( {\bf 
\nabla d} \right) \cdot \left( {\bf d \nabla} \right) + k_{12} \left( 
{\bf\nabla} \cdot {\bf d} \right) \left( {\bf\nabla} \times {\bf d} \right) - 
\cr
& \hskip 0.4cm - k_{13} {\bf\nabla} \left( {\bf\nabla} \cdot {\bf d} \right) +
k_{23} \left[ {\bf\nabla} \left( {\bf d} \cdot {\bf\nabla} \times {\bf d} 
\right) + {\bf d} \cdot {\bf\nabla} \left( {\bf\nabla} \times {\bf d} \right) 
\right] & (A16) \cr} $$
With respect to the local reference frame, it is expressed as
$$ \eqalignno{ {\bf g}^e = & k'_2 \varepsilon^{\alpha\beta} \left( d^\sigma
b_{\sigma\beta} {\bf e}_\alpha - \partial_\beta d_\alpha {\bf e}_3 \right) - 
\cr
& - \left( k'_{33} + 2k'_{13} \right) \left( 2Hb^{\alpha\beta} d_\beta - 
K d^\alpha  \right) {\bf e}_\alpha - \cr
& - k'_{12} 2H \varepsilon^{\alpha\beta} \left( d^\sigma b_{\sigma\beta}
{\bf e}_\alpha - \partial_\beta d_\alpha {\bf e}_3 \right) + \cr
& + k'_{13} a^{\alpha\beta} \partial_\beta \left( 2H - \partial_\sigma
d^\sigma - {1\over 2a} d^\sigma \partial_\sigma a \right) {\bf e}_\alpha + \cr
& + k'_{23} a^{\alpha\beta} \partial_\beta \left( \varepsilon^{\lambda\mu}
\partial_\lambda d_\mu \right) {\bf e}_{\alpha} + O \left( {d^\alpha}^2 
\right) & (A17) \cr} $$

\smallskip
\noindent{{\sl C1.2.2. Viscous resistance}
The rate of strain $ \bf S $ given in Eq.(14) is written for 
a two-dimensional fluid} 
$$ \eqalignno{ \bf S & = {1\over 2} \left( \bf v \bf\nabla_s + \bf\nabla_s 
\bf v \right) \cr
& = {1\over 2} \left( \partial_\beta \bf v \bf e^\beta + \bf e^\alpha 
\partial_\alpha \bf v \right) \cr
& = {1\over 2} \rm g_{\rm ij} \left( t^{\rm j}_\alpha \rm v^{\rm i}_{\;,\,
\beta} + t^{\rm j}_\beta \rm v^{\rm i}_{\;,\,\alpha} \right) \bf e^\alpha \bf 
e^\beta \cr
& = {1\over 2} {\rm g}_{\rm ij} \left[ t^{\rm j}_\alpha \left( t^{\rm i}_\mu 
v^\mu + \dot {\rm r}^{\rm i} \right)_{,\,\beta} + t^{\rm j}_\beta \left( 
t^{\rm i}_\nu v^\nu + \dot {\rm r}^{\rm i} \right)_{,\,\alpha} \right]
{\bf e}^\alpha {\bf e}^\beta \cr
& = {1\over 2} \left( v_{\alpha,\,\beta} + v_{\beta,\,\alpha} - 2wb_{\alpha
\beta} \right) \bf e^\alpha \bf e^\beta & (A18) \cr} $$
for [28] 
$$ {\rm g}_{\rm ij} t^{\rm i}_\lambda t^{\rm j}_\mu = 
a_{\lambda\mu}, \quad {\rm g}_{\rm ij} \left( t^{\rm j}_\lambda 
\dot {\rm r}_{,\,\mu} + t^{\rm j}_\mu \dot {\rm r}^{\rm i}_{,\,\lambda} 
\right) = \dot a_{\lambda\mu} $$
and [38] 
$$ \dot a_{\lambda\mu} = - 2w b_{\lambda\mu} \eqno (71) $$

The rotational velocity of the director $ \bf N $ defined 
in Eq.(15) is written as
$$ \eqalignno{ {\bf N} & = \dot {\bf d} - {1\over 2} {\bf\nabla}_s \times 
\left( {\bf v} - \dot {\bf r} \right) \times {\bf d} \cr
& = \left[ \dot d^\alpha - {1\over 2} v^\beta b^\alpha_\beta + {1\over 2}
\varepsilon^{\lambda\mu} \partial _\lambda v_\mu d_\beta \varepsilon^{\beta
\alpha} \right] {\bf e}_\alpha + \cr
& \hskip 0.4cm + {1\over 2} v^\beta b_{\beta\alpha} d^\alpha {\bf e}_3 + O 
\left( {d^\alpha}^2 \right) & (A19) \cr} $$
where
$$ \dot d^\alpha = \partial_t d^\alpha + v^\sigma 
d^\alpha_{\;,\,\sigma(\tau)}. $$

Inserting Eqs.(A18) and (A19) into relation (25), one obtains
$$ \eqalignno{
{\bf g}^v = 
& \gamma_1 \left( \dot d^\alpha - {1\over 2} v_\beta b^{\beta\alpha}
- {1\over 2} \varepsilon^{\lambda\mu} \partial_\lambda v_\mu d_\beta
\varepsilon^{\beta\alpha} \right) {\bf e}_\alpha - \gamma'_2 2Hw d^\alpha 
{\bf e}_\alpha + \cr
& + \gamma''_2 \left[ {1\over 2} E^{\alpha\beta\lambda\mu} \left( v_{\lambda,
\,\mu} + v_{\mu,\,\lambda} \right) - 2w \left( b^{\beta\alpha} - H a^{\beta
\alpha} \right) \right] d_\beta {\bf e}_\alpha + \cr
& + {1\over 2} \gamma_1 v^\alpha b_{\alpha\beta} d^\beta {\bf e}_3 +
O \left( {d^\alpha}^2 \right) & (A20) \cr} $$
where, $ \gamma'_2 $ is referred to the area dilation, $ \gamma''_2 $ to 
the shear motion, the fourth-order isotropic tensor 
$ E^{\alpha\beta\lambda
\mu} = a^{\alpha\lambda} a^{\beta\mu} + a^{\alpha\mu} a^{\beta\lambda} - 
a^{\lambda\mu} a^{\alpha\beta} $ is used for the tensor operation compatible
with the transverse isotropy of the material [28]. 

\vfill\eject

\smallskip
\leftline{\sl C1.3. Extra surface force}
The substitution of Eq.(A12) into relation (27) gives
$$ \eqalignno{ {\bf\Pi} = & {\partial W\over \partial \left( {\bf d\nabla} 
\right) } \cr
= & k_1 {\bf I} + k_2 {\bf d} \cdot {\bf\varepsilon} + k_{11} \left( 
{\bf\nabla} \cdot {\bf d} \right) {\bf I} - k_{22} \left( {\bf d} \cdot 
{\bf\nabla} \times {\bf d} \right) {\bf d} \cdot {\bf\varepsilon} + \cr
& + \left( k_{33} + 2k_{13} \right) {\bf d} \cdot \left( {\bf\nabla d} \right)
{\bf d} - 2 \left( k_{22} + k_{24} \right) \left[ \left( {\bf\nabla} \cdot 
{\bf d} \right) {\bf I} - {\bf\nabla d} \right] + \cr
& + k_{12} \left[ \left( {\bf\nabla} \cdot {\bf d} \right) {\bf d} \cdot
{\bf\varepsilon} - \left( {\bf d} \cdot {\bf\nabla} \times {\bf d} \right)
{\bf I} \right] - \cr
& - k_{23} \left[ \left( {\bf\nabla} \times {\bf d} \right) {\bf d} - 
{\bf d} \cdot \left( {\bf\nabla d} \right) \cdot {\bf\varepsilon} \right] 
& (A21) \cr} $$
where $ \bf\varepsilon $ is the third-order permutation tensor. With respect 
to the local bases it is given by 
$$ \eqalignno{ {\bf\Pi} = 
& k'_1 a^{\beta\alpha} {\bf e}_\beta {\bf e}_\alpha
+ k'_2 \varepsilon^{\beta\alpha} \left( {\bf e}_\beta {\bf e}_\alpha - 
d_\alpha {\bf e}_\beta {\bf e}_3 \right) - \cr
& - k'_{11} \left( 2H - \partial_\sigma d^\sigma - {1\over 2a} d^\sigma
\partial_\sigma a \right) a^{\beta\alpha} {\bf e}_\beta {\bf e}_\alpha - \cr
& - k'_{22} \varepsilon^{\lambda\mu} \partial_\lambda d_\mu \varepsilon
^{\beta\alpha} {\bf e}_\beta {\bf e}_\alpha - \left( k'_{33} + 2k'_{13} 
\right) d^\sigma b^\beta_\sigma {\bf e}_\beta {\bf e}_3 + \cr
& + 2 \left( k'_{22} + k'_{24} \right) \biggl\{ \biggl[ 2H a^{\beta\alpha} -
b^{\beta\alpha} - \left( \partial_\sigma d^\sigma + {1\over 2a} d^\sigma 
\partial_\sigma a \right) a^{\beta\alpha} + \biggr.\biggr. \cr
& \hskip 2.7cm \biggl.\biggl. + a^{\beta\gamma} d^\alpha_{\;,\,\gamma(\tau)} 
\biggr] {\bf e}_\beta {\bf e}_\alpha + b^{\beta\alpha} d_\alpha 
{\bf e}_\beta {\bf e}_3 \biggr\} - \cr
& - k'_{12} \biggl\{ \biggl[ 2H \varepsilon^{\beta\alpha} - \left( \partial
_\sigma d^\sigma + {1\over 2a} d^\sigma \partial_\sigma a \right) 
\varepsilon^{\beta\alpha} + \biggr.\biggr. \cr 
& \hskip 1.4cm \biggl.\biggl. + \varepsilon^{\lambda\mu} \partial_\lambda 
d_\mu a^{\beta\alpha} \biggr] {\bf e}_\beta {\bf e}_\alpha + 2H \varepsilon^
{\alpha\beta} d_\alpha {\bf e}_\beta {\bf e}_3 \biggr\} - \cr
& - k'_{23} \varepsilon^{\beta\alpha} \partial_\beta d_\alpha
{\bf e}_3 {\bf e}_3 + O \left( {d^\alpha}^2 \right) & (A22) \cr} $$

The application of the surface divergence operator to Eq.(A22) yields 
$$ \eqalignno{ \Pi^{\beta\alpha}_{\;,\,\beta} = 
& k'_2 \varepsilon^{\beta\sigma} d_\sigma b^\alpha_\beta - 
k'_{11} a^{\beta\alpha} \partial_\beta \left( 2H - \partial_\sigma d^\sigma
- {1\over 2a} d^\sigma \partial_\sigma a \right) - \cr
& - k'_{22} \varepsilon^{\beta\alpha} \partial_\beta \left( \varepsilon^
{\lambda\mu} \partial_\lambda d_\mu \right) + 
\left( k'_{33} + 2k'_{13} \right) \left( 2H d^\beta b^\alpha_\beta - K
d^\alpha \right) + \cr
& + 2 \left( k'_{22} + k'_{24} \right) \biggl[ Kd^\alpha - 2H b^{\beta\alpha} 
d_\beta - a^{\beta\alpha} \partial_\beta \left( \partial_\sigma d^\sigma
+ {1\over 2a} d^\sigma \partial_\sigma a \right) + \biggr. \cr
& \hskip 2.4cm \biggl. + a^{\beta\gamma} d^\alpha_{\;,\,\gamma\beta(\tau)} 
\biggr] - \cr
& - k'_{12} \biggl[ \varepsilon^{\beta\alpha} \partial_\beta \left( 2H - 
\partial_\sigma d^\sigma - {1\over 2a} d^\sigma \partial_\sigma a \right) 
+ \biggr. \cr
& \hskip 1.2cm \biggl. + a^{\beta\alpha} \partial_\beta \left( 
\varepsilon^{\lambda\mu} \partial_\lambda d_\mu \right) + 2H \varepsilon
^{\beta\sigma} b^\alpha_\beta d_\sigma \biggr] + 
O \left( {d^\alpha}^2 \right) & (A23) \cr} $$

$$ \eqalignno{ \Pi^{\beta 3}_{\;,\,\beta} = 
& k'_1 2H - k'_2 \varepsilon^{\beta\alpha} \partial_\beta d_\alpha -
k'_{11} \left[ 4H^2 - 2H \left( \partial_\sigma d^\sigma + {1\over 2a} 
d^\sigma \partial_\sigma a \right) \right] - \cr
& -\left( k'_{33} + 2k'_{13} \right) \left[ b^\beta_\sigma d^\sigma
_{\;,\,\beta(\tau)} + d^\beta \partial_\beta \left( 2H \right) \right] + \cr
& + 4 \left( k'_{22} + k'_{24} \right) \biggl[ K - H \left( \partial_\sigma
d^\sigma + {1\over 2a} d^\sigma \partial_\sigma a \right) + \biggr.\cr
& \hskip 2.6cm \biggl. + d^\beta \partial_\beta H + b^{\beta\sigma} 
d_{\sigma,\,\beta(\tau)} \biggr] - \cr
& + k'_{12} \varepsilon^{\beta\alpha} d_\alpha \partial_\beta \left( 2H 
\right) + k'_{23} 2H \varepsilon^{\beta\alpha} \partial_\beta d_\alpha +
O \left( {d^\alpha}^2 \right) & (A24) \cr} $$

\medskip
\leftline{\sl C2. Forces exerting on the continuum}
\smallskip
\leftline{\sl C2.1. Gravitational attraction}
The gravity given in Eq.(4) is written for the two-dimensional fluid
$$ {\bf F}^g = \gamma \hat {\bf g} = \gamma \hat {\bf g}_{\rm k} \left( 
t^{\rm k}_\alpha {\bf e}^\alpha + {\rm n}^{\rm k} {\bf e}^3 
\right) \eqno (A25) $$ 

\smallskip
\leftline{\sl C2.2. Ponderomotive forces} 
The force the external magnetic field exerts on the body of the polar 
fluid, given by Eq.(5), takes the form
$$ {\bf F}^m = {\rm H}_{\rm k,\,j} \left[ \chi_\perp \left( {\rm H}^{\rm j}
- {\rm n}^{\rm j} {\rm H}_{(\rm n)} \right) + \chi_a {\rm H}_{(\rm n)} 
t^{\rm j}_\beta d^\beta \right] \left( t^{\rm k}_\alpha {\bf e}^\alpha + 
{\rm n}^{\rm k} {\bf e}_3 \right) \eqno (A26) $$
In deducing (A26), one used relation (6) and the formula [28] 
$$ t^{\rm i}_\alpha t^{\rm j}_\beta a^{\alpha\beta} = {\rm g}^{\rm ij} -
{\rm n}^{\rm i} {\rm n}^{\rm j} \eqno (A27) $$

The Lorentz force in an electric field (7) is given by 
$$ {\bf F}^e = \gamma_e {\bf E} = \gamma_e \left( {\rm E}_{\rm k}
t^{\rm k}_\alpha {\bf e}^\alpha + E_{(n)} {\bf e}^3 \right) 
\eqno (A28) $$

\smallskip
\leftline{\sl C2.3. Isotropic force}
The application of the surface divergence operator to the isotropic stress
expressed in Eq.(9) gives
$$ {\bf\nabla}_s \cdot {\bf T}^i = - a^{\alpha\beta} \partial_\beta 
\sigma {\bf e}_\alpha \eqno (A29) $$

\smallskip
\leftline{\sl C2.4. Elastic restoring force}
Inserting Eq.(A12) into relation (10) yields
$$ \eqalignno{ {\bf T}^e = & - \left( {\bf\nabla d} \right) \cdot {\partial
W^d \over \partial \left( {\bf d\nabla} \right) } \cr
= & - k_1 {\bf \nabla d} + k_2 \left( {\bf \nabla d} \right) \times {\bf d} 
- k_{11} \left( {\bf\nabla} \cdot {\bf d} \right) \left( {\bf \nabla d} 
\right) - \cr
& - k_{22} \left( {\bf d} \cdot {\bf\nabla} \times {\bf d} \right) \left(
{\bf\nabla d} \right) \times {\bf d} - \cr
& - \left( k_{33} + 2k_{13} \right) \left( {\bf\nabla d} \right) \cdot 
\left( {\bf d\nabla} \right) \cdot \left( {\bf dd} \right) + \cr
& + 2 \left( k_{22} + k_{24} \right) \left[ \left( {\bf\nabla} \cdot {\bf d} 
\right) {\bf\nabla d} - \left( {\bf\nabla d} \right) \cdot \left( {\bf\nabla 
d} \right) \right] + \cr
& + k_{12} \left[ \left( {\bf\nabla} \cdot {\bf d} \right) 
\left( {\bf\nabla d} \right) \times {\bf d} + \left( {\bf d} \cdot 
{\bf\nabla} \times {\bf d} \right) {\bf\nabla d} \right] + \cr 
& + k_{23} \left[ \left( {\bf\nabla d} \right) \times \left( {\bf d \nabla} 
\right) \cdot {\bf d} + \left( {\bf\nabla d} \right) \cdot \left( {\bf\nabla} 
\times {\bf d} \right) {\bf d} \right] & (A30) \cr} $$

With respect to the local bases it is given by 
$$ \eqalignno { {\bf T}^e = 
& k'_1 \left[ \left( b^{\beta\alpha} - a^{\beta\gamma} d^\alpha_
{\;,\,\gamma(\tau)} \right) {\bf e}_\beta {\bf e}_\alpha - 
d_\sigma b^{\sigma\beta} {\bf e}_\beta {\bf e}_3 \right] + \cr
& + k'_2 \varepsilon^{\sigma\alpha} \left[ \left( b^\beta_\sigma - 
a^{\beta\gamma} d_{\sigma,\,\gamma(\tau)} \right) {\bf e}_\beta 
{\bf e}_\alpha + d_\sigma b^\beta_\alpha {\bf e}_\beta {\bf e}_3 \right] - \cr
& - k'_{11} \biggl\{ \left[ 2H b^{\beta\alpha} - 
{1\over\sqrt a} \partial_\sigma \left( d^\sigma \sqrt a \right) 
b^{\beta\alpha} - 2H a^{\beta\sigma} d^\alpha_{\;,\,\sigma(\tau)} \right]
{\bf e}_\beta {\bf e}_\alpha - \biggr. \cr
& \biggl. \hskip 1.2cm - 2H b^{\beta\sigma} d_\sigma {\bf e}_\beta {\bf e}_3 
\biggr\} + k'_{22} \varepsilon^{\lambda\mu} \partial_\lambda d_\mu \varepsilon^ 
{\alpha\sigma} b^\beta_\sigma {\bf e}_\beta {\bf e}_\alpha - \cr
& - \left( k'_{33} + 2k'_{13} \right) \left( 2Hb^{\beta\alpha} d_\alpha - K 
d^\beta \right) {\bf e}_\beta {\bf e}_3 + \cr
& + 2 \left( k'_{22} + k'_{24} \right) \biggl\{ \biggl[ K a^{\beta\alpha} + 
a^{\beta\gamma} d_{\sigma,\,\gamma(\tau)} b^{\sigma\alpha} - 
{1\over\sqrt a} \partial_\sigma \left( d^\sigma \sqrt a \right)
b^{\beta\alpha} + \biggr.\biggr. \cr
& \hskip 2.4cm \biggl.\biggl. + \left( b^{\sigma\beta} - 2H a^{\sigma\beta} 
\right) d^\alpha_{\;,\,\sigma(\tau)} \biggr] {\bf e}_\beta {\bf e}_\alpha - 
K d^\beta {\bf e}_\beta {\bf e}_3 \biggr\} - \cr
& - k'_{12} \biggl\{ \biggl[ \left( 2H b^\beta_\mu - 
{1\over\sqrt a} \partial_\sigma \left( d^\sigma \sqrt a \right) b^\beta_\mu 
- 2H a^{\beta\sigma} d_{\mu,\,\sigma(\tau)} \right) 
\varepsilon^{\mu\alpha} + \biggr.\biggr. \cr 
& \biggl.\biggl. \hskip 1.4cm + \varepsilon^{\lambda\mu} \partial_\lambda 
d_\mu b^{\alpha\beta} \biggr] {\bf e}_\beta {\bf e}_\alpha - 2H \varepsilon
^{\lambda\mu} b^\beta_\lambda d_\mu {\bf e}_\beta {\bf e}_3 \biggr\} - \cr
& - k'_{23} K \varepsilon^{\beta\alpha} d_\alpha {\bf e}_\beta {\bf e}_3 
+ O \left( {d^\alpha}^2 \right) & (A31) \cr} $$

The dot product of the surface derivative operator and the stress 
$ {\bf T}^e $ gives
$$ \eqalignno{ {T^e}^{\beta\alpha}_{\;,\,\beta} = 
& k'_1 \left[ a^{\beta\alpha} \partial_\beta \left( 2H \right) -
d^\alpha_{\;,\,\sigma\beta(\tau)} a^{\beta\sigma} + 2H b^{\beta\alpha} 
d_\beta - K d^\alpha \right] - \cr 
& - k'_2 \left\{ \varepsilon^{\alpha\beta} \left[ 
\partial_\beta \left( 2H \right) + K d_\beta - a^{\sigma\gamma} d_{\beta,
\,\gamma\sigma(\tau)} \right] 
- 2H \varepsilon^{\lambda\mu} b^\alpha_\lambda d_\mu \right\} - \cr
& - k'_{11} \biggl\{ \left( b^{\beta\alpha} + 2H a^{\beta\alpha} \right) 
\partial_\beta \left( 2H \right) - a^{\beta\sigma} d^\alpha
_{\;,\,\sigma(\tau)} \partial_\beta \left( 2H \right) - \biggr. \cr 
& \biggl. \hskip 1.2cm - \left( b^{\beta\alpha} \partial_\beta + 
2a^{\beta\alpha} \partial_\beta H \right) \left[ {1\over\sqrt a} 
\partial_\sigma \left( d^\sigma \sqrt a \right) \right] + \biggr. \cr 
& \biggl. \hskip 1.2cm + 2H \left( 2H b^{\beta\alpha} d_\beta - a^{\beta
\sigma} d^\alpha_{\;,\,\sigma\beta(\tau)} - K d^\alpha \right) \biggr\} + \cr
& + k'_{22} \left[ \varepsilon^{\alpha\sigma} b^\beta_\sigma \partial_\beta
\left( \varepsilon^{\lambda\mu} \partial_\lambda d_\mu \right) + 
\varepsilon^{\alpha\beta} \partial_\beta \left( 2H \right) 
\left( \varepsilon
^{\lambda\mu} \partial_\lambda d_\mu \right) \right] + \cr
& + \left( k'_{33} + 2k'_{13} \right) \left( 4H^2 b^{\beta\alpha} d_\beta - 
2HK d^\alpha - K b^{\alpha\beta} d_\beta \right) + \cr
& + 2 \left( k'_{22} + k'_{24} \right) 
\biggl\{ a^{\beta\alpha} \partial_\beta 
K + K b^{\beta\alpha} d_\beta + a^{\beta\sigma} \left( b^{\alpha\gamma} 
d_{\gamma,\,\sigma(\tau)} \right)_{,\,\beta(\tau)} + \biggr. \cr 
& \hskip 2.4cm \biggl. + \left( b^{\beta\sigma} - 2H a^{\beta\sigma} \right) 
d^\alpha_{\;,\,\sigma\beta(\tau)} - \biggr. \cr 
& \hskip 2.4cm \biggl. - \left( b^{\alpha\beta} \partial_\beta + 
2a^{\alpha\beta} \partial_\beta H \right) \left[ {1\over\sqrt a} 
\partial_\sigma \left( \sqrt a d^\sigma \right) \right] \biggr\} - \cr 
& - k'_{12} \biggl\{ 2\varepsilon^{\beta\alpha} \left( b_\beta^\sigma 
\partial_\sigma H + \partial_\beta H^2 \right) + 4H^2 \varepsilon^{ \beta 
\sigma} b^\alpha_\sigma d_\beta - \biggr. \cr 
& \hskip 1.2cm \biggl. - \varepsilon^{\beta\alpha} \biggl[ \left( 2\partial 
_\beta H + b^\sigma_\beta \partial_\sigma \right) 
\left( {1\over\sqrt a} \partial_\gamma \left( \sqrt a d^\gamma \right) 
\right) + \biggr.\biggr. \cr 
& \hskip 1.2cm \biggl.\biggl. + 2a^{\gamma\sigma} \partial_\gamma H 
d_{\beta,\,\sigma(\tau)} + 2H a^{\gamma\sigma} d_{\beta,\,\sigma\gamma 
(\tau)} - 2HK d_\beta \biggr] - \biggr. \cr 
& \hskip 1.2cm \biggl. - \left( b^{\beta\alpha} \partial_\beta + 
2a^{\beta\alpha} \partial_\beta H \right) \left( \varepsilon^{\lambda\mu} 
\partial_\lambda d_\mu \right) \biggr\} + \cr 
& + k'_{23} K b^\alpha_\beta \varepsilon^{\beta\gamma} d_\gamma + 
O \left( {d^\alpha}^2 \right) & (A32) \cr} $$ 

$$ \eqalignno{ {T^e}^{\beta 3}_{\;,\,\beta} = 
& k'_1 \left[ 4H^2 - 2K - 2b^\beta_\sigma d^\sigma_{\;,\,\beta(\tau)} -
d^\sigma \partial_\sigma \left( 2H \right) \right] + k'_2 \varepsilon
^{\beta\alpha} d_\beta \partial_\alpha \left( 2H \right) - \cr
& -k'_{11} \biggl[ \left( 4H^2 - 2K \right) \left( 2H - \partial_\sigma 
d^\sigma - {1\over 2a} d^\sigma \partial_\sigma a \right) - \biggr. \cr
& \hskip 1.2cm \biggl. - \left( d^\sigma b^\beta_\sigma + 2H d^\beta \right) 
\partial_\beta \left( 2H \right) - 4H b^\beta_\sigma d^\sigma
_{\;,\,\beta(\tau)} \biggr] - \cr
& - \left( k'_{33} + 2k'_{13} \right) \biggl[ 2H b^\beta_\sigma 
d^\sigma_{\;,\,\beta(\tau)} + \left( b^\beta_\sigma d^\sigma + 2H d^\beta
\right) \partial_\beta \left( 2H \right) - \biggr. \cr
& \hskip 2.5cm \biggl. - d^\beta \partial_\beta K - K \left( \partial_\sigma 
d^\sigma + {1\over 2a} d^\beta \partial_\beta a \right) \biggr] + \cr
& + 2 \left( k'_{22} + k'_{24} \right) 
\biggl[ 2HK - d^\beta \partial_\beta K
+ 2H b^{\beta\alpha} d_{\alpha,\,\beta(\tau)} - \biggr. \cr
& \hskip 2.4cm  \biggl. - \left( 4H^2 + K \right) \left( \partial_\sigma
d^\sigma + {1\over 2a} d^\sigma \partial_\sigma a \right) \biggr] + \cr
& + 2K'_{12} \biggl\{ \varepsilon^{\alpha\beta} \left[ \left( b^\sigma_\alpha
\partial_\sigma H + \partial_\alpha H^2 \right) d_\beta + H b^\beta_\alpha
d_{\sigma,\,\beta(\tau)} \right] - \biggr. \cr
& \hskip 1.5cm \biggl. - \left( 2H^2 - K \right) \varepsilon^{\lambda\mu} 
\partial_\lambda d_\mu \biggr\} + \cr
& + k'_{23} \varepsilon^{\alpha\beta} \left( 
d_\alpha \partial_\beta K + K d_{\alpha,\,\beta(\tau)} \right) + 
O \left( {d^\alpha}^2 \right) & (A33) \cr} $$

\smallskip
\leftline{\sl C2.5. Viscous resistance}
With the help of Eqs.(A18)-(A19), the viscous stress Eq.(12) is written as
$$ \eqalignno{ {\bf T}^v = 
& - \biggl\{ 2\mu Hw a^{\alpha\beta} + 2\eta w \left( b^{\alpha\beta} - 
H a^{\alpha\beta} \right) - {\eta\over 2} E^{\alpha\beta\lambda\mu} 
\left( v_{\lambda,\,\mu} + v_{\mu,\,\lambda} \right) + \biggr. \cr 
& \hskip 0.5cm \biggl. + {\alpha_2\over 2} v^\sigma b^\beta_\sigma d^\alpha + 
{\alpha_3\over 2} v^\sigma b^\alpha_\sigma d^\beta \biggr\} 
{\bf e}_\alpha {\bf e}_\beta + \cr 
& + \left\{ \alpha_3 \left[ \dot d^\alpha - {1\over 2} v^\beta 
b^\alpha_\beta - {1\over 2} \left( \varepsilon^{\mu\lambda} v_{\lambda,\,\mu}
\right) \varepsilon^{\beta\alpha} d_\beta \right] - \alpha'_6 2Hw d^\alpha + 
\right. \cr 
& \hskip 0.5cm \left. + \alpha''_6 \left[ {1\over 2} 
E^{\alpha\beta\lambda\mu} 
\left( v_{\lambda,\,\mu} + v_{\mu,\,\lambda} \right) d_\beta - 
2w \left( b^{\alpha\beta} d_\beta - H d^\alpha \right) \right] \right\} 
{\bf e}_\alpha {\bf e}_3 + \cr 
& + \left\{ \alpha_2 \left[ \dot d^\beta - {1\over 2} v^\alpha 
b^\beta_\alpha - {1\over 2} \left( \varepsilon^{\mu\lambda} v_{\lambda,\,\mu} 
\right) \varepsilon^{\alpha\beta} d_\alpha \right] - \alpha'_5 2Hw d^\beta + 
\right. \cr 
& \hskip 0.5cm \left. + \alpha''_5 \left[ {1\over 2} 
E^{\alpha\beta\lambda\mu} \left( v_{\lambda,\,\mu} + v_{\mu,\,\lambda} 
\right) d_\alpha - 2w \left( d_\alpha b^{\alpha\beta} - H d^\beta \right) 
\right] \right\} {\bf e}_3 {\bf e}_\beta + \cr 
& + {1\over 2} \left( \alpha_2 + \alpha_3 \right) v_\alpha b^{\alpha\beta} 
d_\beta {\bf e}_3 {\bf e}_3 + O \left( {d^\alpha}^2 \right) & (A34) \cr} $$ 
where the viscosity coefficient $ \alpha_4 $ in Eq.(12) has been split 
into the dilatation viscosity $ \mu $ and the shear viscosity $ \eta $, 
the $ \alpha' $ coefficients refer to the dilatation of area and 
the $ \alpha'' $ coefficients to the shear motion. 

The dot product of $ {\bf\nabla}_s $ and $ {\bf T}^v $ gives
$$ \eqalignno{ {T^v}^{\beta\alpha}_{\;,\,\beta} = 
& - \mu a^{\alpha\beta} \partial_\beta \left( 2Hw \right) + {\eta\over 2} 
E^{\beta\alpha\lambda\mu} \left( v_{\lambda,\,\mu\beta} + 
v_{\mu,\,\lambda\beta} \right) - \cr
& - 2\eta \left[ \partial_\beta w \left( b^{\beta\alpha} - H a^{\beta\alpha}
\right) + w a^{\alpha\beta} \partial_\beta H \right] + \cr
& + \alpha_2 H v_\sigma b^{\sigma\alpha} + {\alpha_3\over 2} \left( 2H 
b^{\alpha\sigma} v_\sigma - K v^\alpha \right) - \cr
& - \alpha_2 \left[ 2H \dot d^\alpha + {1\over 2} \left( v_\sigma
b^{\sigma\alpha} d^\beta \right)_{,\,\beta(\tau)} - H \left( \varepsilon^
{\mu\lambda} v_{\lambda,\,\mu} \right) \varepsilon^{\beta\alpha} b_\beta
\right] - \cr
& - \alpha_3 \left[ \dot d^\beta b^\alpha_\beta + {1\over 2} 
b^{\beta\sigma} \left( v_\sigma d^\alpha \right)_{,\,\beta(\tau)} + d^\alpha
v^\sigma \partial_\sigma H - \right. \cr
& \hskip 1cm \left. - {1\over 2} \left( \varepsilon^{\mu\lambda} 
v_{\lambda,\,\mu} \right) \varepsilon^{\sigma\beta} d_\sigma b^\alpha_\beta 
\right] + \cr
& + \alpha'_5 4H^2 w d^\alpha + \alpha'_6 2Hw b^{\alpha\beta} 
d_\beta - \cr
& - \alpha''_5 \left[ HE^{\alpha\beta\lambda\mu} \left( v_{\lambda,\,\mu} +
v_{\mu,\,\lambda} \right) d_\beta - 4Hw \left( d_\beta b^{\beta\alpha} - H
d^\alpha \right) \right] - \cr
& - \alpha''_6 \left[ {1\over 2} E^{\alpha\beta\lambda\mu} 
\left( v_{\lambda,\,\mu} + v_{\mu,\,\lambda} \right) d_\sigma b^\alpha_\beta 
- 2w \left( H b^{\alpha\sigma} d_\sigma - Kd^\alpha \right) \right] + \cr
& + O \left( {d^\alpha}^2 \right) & (A35) \cr} $$

$$ \eqalignno{ {T^v}^{\beta 3}_{\;,\,\beta(\tau)} = 
& - \mu 4H^2w + \eta \left( b^{\lambda\mu} - H a^{\lambda\mu} \right) 
\left( v_{\lambda,\,\mu} + v_{\mu,\,\lambda} \right) - \cr
& - 4\eta w \left( H^2 - K \right) - {\alpha_3\over 2} \left( b^\beta_\sigma
v^\sigma_{\;,\,\beta} + 2 v^\beta \partial_\beta H \right) + \cr
& + \alpha_3 \biggl\{ \partial_t \left( \partial_\beta d^\beta + 
{1\over 2a} d^\beta \partial_\beta a \right) + 
v^\sigma_{\;,\,\beta} d^\beta_{\;,\,\sigma(\tau)} + \biggr. \cr
& \hskip 1.1cm \biggl. + v^\sigma d^\beta_{\;,\,\sigma\beta(\tau)} 
- {1\over 2} \varepsilon^{\beta\sigma} \left( d_\sigma \partial_\beta + 
\partial_\beta d_\sigma \right) \left( \varepsilon^{\lambda\mu} 
\partial_\lambda v_\mu \right) \biggr\} - \cr
& - \alpha'_6 \left[ d^\beta \partial_\beta \left( 2Hw \right) + 2Hw 
\left( \partial_\beta d^\beta + {1\over 2a} d^\beta \partial_\beta a \right) 
\right] + \cr
& + \alpha''_6 \biggl\{ {1\over 2} E^{\alpha\beta\lambda\mu} \left[ \left(
v_{\mu,\,\lambda} + v_{\lambda,\,\mu} \right) d_\alpha \right]_{,\,\beta
(\tau)} - \biggr. \cr
& \hskip 1.2cm \biggl. - 2 \partial_\beta w \left( d_\sigma b^{\sigma\beta} -
H d^\beta \right) - 2w \biggl[ d_{\alpha,\,\beta(\tau)} b^{\beta\alpha} +
\biggr.\biggr. \cr
& \hskip 1.5cm \biggl.\biggl. + d^\beta \partial_\beta H - H \left( 
\partial_\beta d^\beta + {1\over 2a} d^\beta \partial_\beta a \right) 
\biggr] \biggr\} - \cr
& - \left( \alpha_2 + \alpha_3 \right) H v^\alpha b_{\alpha\beta} d^\beta
+ O \left( {d^\alpha}^2 \right) & (A36) \cr} $$

\bigskip
\leftline{\bf Appendix D. Energy balance}
\medskip

In the following, one looks for a testable parameter to alter the internal 
energy $ U $ in the energy balance equation.

With the help of Eqs.(8), (9) and (23), Eq.(97) is written as
$$ \eqalignno{ & \gamma \dot U - \left( {T^e}^{\beta\alpha} v_{\alpha,\,
\beta} + {T^e}^{\beta 3} \partial_\beta w + \Pi^{\beta\alpha} {\dot d}
_{\alpha,\,\beta} - {g^e}^\alpha \dot d_\alpha \right) \cr 
& \hskip 0.5cm = Q^h - {J^h}^\alpha_{\;,\,\alpha} + \left[ \left( - \sigma 
a^{\beta\alpha} + {T^v}^{\beta\alpha} \right) v_{\alpha,\,\beta} + 
{T^v}^{\beta 3} \partial_\beta w - {g^v}^\alpha \dot d_\alpha \right] 
& (A37) \cr } $$

From the first law of thermodynamics, one knows that
$$ dU - \sum_i L_i dl_i = dQ + dW_{diss} \eqno (A38) $$
where $ W_{diss} $ is the dissipated work, $ l_i $ the $ ith $ 
work coordinate, $ L_i $ the conjugate work coefficient, 
and $ U $ and $ Q $ are defined as usual.

As $ U $ is the characteristic function of the temperature 
$ \Theta $ and of 
the work coefficient $l_i $, one has
$$ dU = \left( {\partial U \over \partial \Theta} \right)_{l_i} d\Theta +
\sum_i \left( {\partial U \over \partial l_i} \right)_{\Theta,\,l_j} d l_i. 
$$ 
Equality (A38) may thus be written as
$$ \left( {\partial U \over \partial \Theta} \right)_{l_i} d \Theta + \sum_i
\left[ \left( {\partial U \over \partial {l_i} } \right)_{\Theta,\,j} - L_i
\right] d l_{l_i} = dQ + dW_{diss} \eqno (A39) $$
From the fundamental equations
$$ \eqalign{ dU & = \Theta dS + \sum_i L_i dl_i \cr
dF & = -S d\Theta + \sum_i L_i dl_i, \cr} $$
one has
$$ \eqalign{ \left( {\partial U \over \partial {l_i} } \right)_{l_j} & 
= \Theta \left( {\partial S \over \partial {l_i} } \right)_{l_j} + L_i \cr
\left( {\partial S \over \partial {l_i} } \right)_{\Theta,\,l_j} & =
- \left( {\partial L_i \over \partial \Theta } \right)_{l_i}, \cr} $$
where $ F $ is the Helmholtz function and $ S $ is the entropy. 
Substitution of 
the two expressions above into Eq.(A39) leads to
$$ C_l d\Theta - \Theta \sum_i \left( {\partial L_i \over \partial \Theta}
\right)_{l_i} dl_i = dQ + dW_{diss} \eqno (A40) $$
where
$$ C_l \equiv \left( {\partial U \over \partial \Theta } \right)_{l_i}
= \Theta \left( {\partial S \over \partial \Theta } \right)_{l_i} 
\eqno (A41) $$
is defined as the heat capacity at constant work coordinates [39]. 
Dividing Eq.(A40) 
by the infinitesimal time interval, one obtains the instantaneous 
energy balance
$$ C_l {d\Theta \over dt} - \Theta \sum_i \left( {\partial L_i \over 
\partial \Theta} \right)_{l_i} { dl_i \over dt} = {dQ\over dt} + 
{dW_{diss}\over dt} \eqno (A42) $$

We turn back now to Eq.(A37). The terms in the parentheses on the 
left-side of Eq.(A37), i.e. the rate of change of the work of
the elastic deformation, 
is identical to the second term on the left-side of Eq.(A42), 
whereas the right-sides of Eqs.(A37) and (A42) are identical. It follows 
thus that
$$ \eqalignno{ & \gamma C_l \dot\Theta - {T^e}^{\beta\alpha} v_{\alpha,\,
\beta} - {T^e}^{\beta 3} \partial_\beta w - \Pi^{\beta\alpha} {\dot d}
_{\alpha,\,\beta} + {g^e}^\alpha \dot d_\alpha \cr 
& \hskip 0.5cm = Q^h - {J^h}^\alpha_{\;,\,\alpha} + \left( - \sigma 
a^{\beta\alpha} + {T^v}^{\beta\alpha} \right) v_{\alpha,\,\beta} + 
{T^v}^{\beta 3} \partial_\beta w - {g^v}^\alpha \dot d_\alpha 
& (A43) \cr } $$
Here the convected derivative of the temperature is given by 
$$ \dot\Theta = \partial_t \Theta + \left( v^\alpha t^{\rm j}_\alpha 
+ w {\rm n}^{\rm j} \right) \partial_{\rm j} \Theta. $$
The internal energy $ U $ in Eq.(A37) has now been replaced by the product
$ C_l \dot\Theta $.

The heat flux Eq.(29) is expanded with respect to the local bases
$$ \eqalignno{ {\bf J}^h = 
& - \partial_{\rm j} \Theta \left\{ \left( \kappa_\perp t^{\rm j}_\alpha + 
\kappa_a {\rm n}^{\rm j} d_\alpha \right) {\bf e}^\alpha + 
\left( \kappa_a t^{\rm j}_\alpha d^\alpha + \kappa_\parallel {\rm n}^{\rm j} 
\right) {\bf e}^3 \right\} + O \left( {d^\alpha}^2 \right) & (A44) \cr} $$
where the coefficient $ \beta_0 $ has been replaced by the heat conductivity
perpendicular to the director vector $ - \kappa_\perp $ and the coefficient
$ \beta_1 $ by the thermal anisotropy $ - \kappa_a \equiv \kappa_\perp -
\kappa_\parallel $.

The divergence of the heat flux is given by 
$$ \eqalignno{ {J^h}^\alpha_{,\,\alpha} = & {J^h}^\alpha_{,\,\alpha(\tau)} -
2H {J^h}^3 \cr
= & - \kappa_\perp \left( \partial_{\rm j} \Theta \right)_{,\,\rm k} 
\left( \rm g^{\rm kj} - \rm n^{\rm k} \rm n^{\rm j} \right) + \kappa_a 2H 
\left( {\rm n}^{\rm j} \partial_{\rm j} \Theta \right) + \cr
& + \kappa_a \biggl[ \left( t^{\rm j}_\beta \partial_{\rm j} \Theta \right) 
\left( 2H d^\beta + b^\beta_\alpha d^\alpha \right) - \biggr. \cr
& \hskip 0.4cm \biggl. \hskip 0.7cm - \left( {\rm n}^{\rm j} \partial_{\rm j} 
\Theta \right) \left( \partial_\alpha + {1\over 2a} \partial_\alpha a \right) 
d^\alpha - \biggr. \cr
& \hskip 1.1cm \biggl. - \rm n^{\rm j} \left( \partial_{\rm j} \Theta \right)
_{,\,\rm k} t^{\rm k}_\alpha d^\alpha \biggr] + O \left( {d^\alpha}^2 \right) 
& (A45) \cr} $$

\vfill\eject

\leftline{\bf References}
\medskip

\item{1.} F. Brochard and J. F. Lennon, J. Phys. (Paris), {\bf 36} (1975) 
1035.
\item{2.} H. C. Pant and B. Rosenberg, Biochim. Biophys. Acta, {\bf 225}
(1971) 379.
\item{3.} R. Larter, Chem. Rev., {\bf 90} (1990) 355.
\item{4.} C. W. Oseen, Trans. Faraday Soc., {\bf 29} (1933) 883.
\item{5.} F. C. Frank, Discuss. Faraday Soc., {\bf 25} (1958) 19.
\item{6.} J. Nehring and A. Saupe, J. Chem. Phys., {\bf 54} (1971) 337.
\item{7.} W. Helfrich, Z. Naturforsch., Teil C, {\bf 28} (1973) 693.
\item{8.} X. Michalet, F. J\" ulicher, B. Fourcade, U. Seifert and D. 
Bensimon, Recherc., {\bf 25} (1994) 1012.
\item{9.} J. L. Ericksen, Kolloid. Z., {\bf 173} (1960) 117; Arch.
Rational Mech. Anal., {\bf 4} (1960) 231, {\bf 9} (1962) 26; 
Trans. Soc. Rheol., {\bf 4} (1960) 29, {\bf 5} (1961) 23; 
Quart. J. Mech. Appl. Math., {\bf 29} (1976) 203.
\item{10.} F. M. Leslie, Quart. J. Mech. Appl. Math., {\bf 19} (1966) 357; 
Arch. Rational Mech. Anal., {\bf 28} (1968) 265; Adv. Liq. Cryst., 
{\bf 4} (1979) 1. 
\item{11.} O. Parodi, J. Phys. (Paris), {\bf 31} (1970) 581.
\item{12.} M. J. Stephen, J. Phys. Rev., {\bf A2} (1970) 1558.
\item{13.} M. J. Stephen and J. P. Straley, Rev. Mod. Phys., {\bf 46} 
(1974) 617.
\item{14.} R. Williams, J. Chem. Phys., {\bf 39} (1963) 384.
\item{15.} W. Greubel and U. Wolff, Appl. Phys. Lett., {\bf 19} 
(1971) 213.
\item{16.} M. F. Schiekel and K. Fahrenschon, Appl. Phys. Lett., {\bf 19} 
(1971) 391.
\item{17.} N. Felici, Rev. Gen. Electr., {\bf 78} (1969) 717.
\item{18.} W. Helfrich, J. Chem. Phys., {\bf 51} (1969) 4092.
\item{19.} P. G. de Gennes, Commun. Solid State Phys., {\bf 3} (1970) 35; 
{\bf 3} (1971) 148. 
\item{20.} J. M. Schneider and P. K. Watson, Phys. Fluids, {\bf 13}
(1970) 1948, 1955. 
\item{21.} E. Dubois-Violette, P. G. Gennes and O. Parodi, J. Phys. 
(Paris), {\bf 32} (1971) 305.
\item{22.} W. J. A. Goossens, Adv. Liq. Cryst., {\bf 3} (1978) 1.
\item{23.} H. M\o llmann, Introduction to the Theory of Thin Shells, John 
Wiley, New York, 1981.
\item{24.} M. A. Peterson, J. Math. Phys., {\bf 26} (1985) 711; 
Mol. Cryst. Liq. Cryst., {\bf 127} (1985) 257.
\item{25.} E. Evans, A. Yeung, R. Waugh and J. Song, in
R. Lipowsky, D. Richter and K. Kremer (Eds.), The Structure and 
Conformation of Amphiphilic Membranes, 
Springer-Verlag, Berlin, 1992, p.148.
\item{26.} U. Seifert and S. A. Langer, Europhys. Lett., {\bf 23} (1993) 71. 
\item{27.} J. S. Dahler and L. E. Scriven, Nature, {\bf 192} (1961) 36.
\item{28.} R. Aris, Vector, Tensor and the Basic Equations of Hydrodynamics, 
Prentice-Hall, Englewood Cliffs, NJ, 1962.
\item{29.} D. Landau and E. M. Lifshitz, Electrodynamics of Continuous Media, 
Pergamon Press, Oxford, 1984.
\item{30.} C. E. Weatherburn, Differential Geometry of Three Dimensions, 
Cambridge University Press, London, 1955.
\item{31.} Zh. C. Ou-Yang and S. Liu, Mol. Cryst. Liq. Cryst., {\bf 204} 
(1991) 143.
\item{32.} N. Nakashima, S. Asakuma and T. Kunitake, J. Am. Chem. Soc., 
{\bf 107} (1985) 509.
\item{33.} N. Nakashima, S. Asakuma, J-M. Kim and T. Kunitake, Chem. 
Lett., {\bf 1984} (1984) 1709.
\item{34.} E. A. Evans and P. L. La Cell, Blood, {\bf 45} (1975) 29.
\item{35.} E. A. Evans, R. Waugh and L. Melnik, Biophys. J., {\bf 16} 
(1976) 585.
\item{36.} E. A. Evans and R. Skalak, Mechanics and Thermodynamics of 
Biomembranes, CRC Press, Boca Raton, Florida, 1980.
\item{37.} A. G. Lee, Prog. Biophys. Molec. Biol., {\bf 29} (1975) 3.
\item{38.} C. Truesdell and R. A. Toupin, The Classical Field Theories, in
S. Flugge (Ed.), Encyclopedia of Physics, Vol.3, No.1, Springer-Verlay, 
Berlin, 1960.
\item{39.} R. Haase, Thermodynamics of Irreversible Processes, Dover 
Publications, Mineola, New York, 1990.

\vskip3cm
\leftline{\sl Figure captions}
\medskip
Fig.1 Illustration of the middle surface and the local coordinate axes.

\end